\documentclass[twocolumn, resetfootnote]{aastex701}

\shorttitle{Sparse by the River}
\shortauthors{Bình et al.}

\usepackage[symbol]{footmisc}
\usepackage[T5]{fontenc}
\usepackage[utf8]{inputenc}
\usepackage{amsmath,amssymb}
\usepackage{subfig}
\usepackage{graphicx}
\usepackage{booktabs}

\begin{document}

\title{Sparse by the River: Diverse Environments of z > 3 Massive Quiescent Galaxies\footnote{The title is translated from ``lác đác bên sông'', a phrase in a topographical poem by nineteenth-century Vietnamese poetess Nguyễn Thị Hinh/Bà Huyện Thanh Quan.}}

\author[0009-0009-9700-1811]{Nguyễn Bình}
\affiliation{Department of Astronomy, University of Washington, Seattle, WA 98195, USA}
\email[show]{ngbinh@uw.edu}

\author[0000-0002-7530-8857]{Arianna S. Long}
\affiliation{Department of Astronomy, University of Washington, Seattle, WA 98195, USA}
\email{aslong@uw.edu}

\author[0000-0002-0243-6575]{Jacqueline Antwi-Danso}
\affiliation{David A. Dunlap Department of Astronomy \& Astrophysics, University of Toronto, 50 St George Street, Toronto, ON M5S 3H4, Canada}
\affiliation{Dunlap Institute for Astronomy and Astrophysics, 50 St. George Street, Toronto, Ontario, M5S 3H4, Canada}
\affiliation{Department of Astronomy, University of Massachusetts, Amherst, MA 01003, USA}
\email{j.antwidanso@utoronto.ca}

\author[0009-0000-2577-1619]{David C. Andrews}
\affiliation{Department of Astronomy, University of Washington, Seattle, WA 98195, USA}
\email{davida04@uw.edu}

\author[0009-0005-3133-1157]{Greta Toni}
\affiliation{University of Bologna - Department of Physics and Astronomy “Augusto Righi” (DIFA), Via Gobetti 93/2, I-40129 Bologna, Italy}
\affiliation{INAF- Osservatorio di Astrofisica e Scienza dello Spazio, Via Gobetti 93/3, I-40129, Bologna, Italy}
\affiliation{Zentrum f\"{u}r Astronomie, Universit\"{a}t Heidelberg, Philosophenweg 12, D-69120, Heidelberg, Germany}
\email{greta.toni4@unibo.it}

\author[0000-0002-6184-9097]{Jaclyn B. Champagne}
\affiliation{Steward Observatory, University of Arizona, 933 N Cherry Ave, Tucson, AZ 85721, USA}
\email{jbchampagne@arizona.edu}

\author[0000-0003-3596-8794]{Hollis B. Akins}
\altaffiliation{NSF Graduate Research Fellow}
\affiliation{The University of Texas at Austin, 2515 Speedway Blvd Stop C1400, Austin, TX 78712, USA}
\email{hollis.akins@gmail.com}

\author[0009-0008-5008-4309]{Tiara Anderson}
\affiliation{Department of Astronomy, University of Washington, Seattle, WA 98195, USA}
\email{tiaraa2@uw.edu}

\author[0000-0002-0569-5222]{Rafael C. Arango-Toro}
\affiliation{Aix Marseille Univ, CNRS, CNES, LAM, Marseille, France}
\email{rafael.arango-toro@lam.fr}

\author[0000-0002-0930-6466]{Caitlin M. Casey}
\affiliation{Department of Physics, University of California, Santa Barbara, Santa Barbara, CA 93106, USA}
\affiliation{Cosmic Dawn Center (DAWN), Denmark}
\email{cmcasey@ucsb.edu}

\author[0000-0001-8551-071X]{Yingjie Cheng}
\affiliation{Department of Astronomy, University of Washington, Seattle, WA 98195, USA}
\email{yingjiec@uw.edu}

\author[0000-0003-3881-1397]{Olivia R. Cooper}
\altaffiliation{NSF Astronomy and Astrophysics Postdoctoral Fellow}
\affiliation{Department for Astrophysical \& Planetary Science, University of Colorado, Boulder, CO 80309, USA}
\email{olivia.cooper@colorado.edu}

\author[0000-0003-4761-2197]{Nicole E. Drakos}
\affiliation{Department of Physics and Astronomy, University of Hawaii, Hilo, 200 W Kawili St, Hilo, HI 96720, USA}
\email{ndrakos@hawaii.edu}

\author[0000-0002-9382-9832]{Andreas L. Faisst}
\affiliation{Caltech/IPAC, MS 314-6, 1200 E. California Blvd. Pasadena, CA 91125, USA}
\email{afaisst@caltech.edu}

\author[0000-0002-3560-8599]{Maximilien Franco}
\affiliation{Université Paris-Saclay, Université Paris Cité, CEA, CNRS, AIM, 91191 Gif-sur-Yvette, France}
\email{maximilien.franco@cea.fr}

\author[0009-0007-0553-9610]{Elaine Gammon}
\affiliation{Department of Astronomy, University of Washington, Seattle, WA 98195, USA}
\email{egammon@uw.edu}

\author[0000-0002-3301-3321]{Michaela Hirschmann}
\affiliation{Institute of Physics, GalSpec, Ecole Polytechnique Federale de Lausanne, Observatoire de Sauverny, Chemin Pegasi 51, 1290 Versoix, Switzerland}
\affiliation{INAF, Astronomical Observatory of Trieste, Via Tiepolo 11, 34131 Trieste, Italy}
\email{michaela.hirschmann@epfl.ch}

\author[0000-0002-7303-4397]{Olivier Ilbert}
\affiliation{Aix Marseille Univ, CNRS, CNES, LAM, Marseille, France}
\email{olivier.ilbert@lam.fr}

\author[0000-0001-9187-3605]{Jeyhan S. Kartaltepe}
\affiliation{Laboratory for Multiwavelength Astrophysics, School of Physics and Astronomy, Rochester Institute of Technology, 84 Lomb Memorial Drive, Rochester, NY 14623, USA}
\email{jeyhan@astro.rit.edu}

\author[0000-0002-6610-2048]{Anton M. Koekemoer}
\affiliation{Space Telescope Science Institute, 3700 San Martin Dr., Baltimore, MD 21218, USA} 
\email{koekemoer@stsci.edu}

\author[0000-0001-9773-7479]{Daizhong Liu}
\affiliation{Purple Mountain Observatory, Chinese Academy of Sciences, 10 Yuanhua Road, Nanjing 210023, China}
\email{dzliu@pmo.ac.cn}

\author[0000-0002-4872-2294]{Georgios E. Magdis}
\affiliation{Cosmic Dawn Center (DAWN), Denmark} 
\affiliation{DTU-Space, Technical University of Denmark, Elektrovej 327, 2800, Kgs. Lyngby, Denmark}
\affiliation{Niels Bohr Institute, University of Copenhagen, Jagtvej 128, DK-2200, Copenhagen, Denmark}
\email{geoma@space.dtu.dk}

\author[0000-0002-3517-2422]{Matteo Maturi}
\affiliation{Zentrum f\"{u}r Astronomie, Universit\"{a}t Heidelberg, Philosophenweg 12, D-69120, Heidelberg, Germany}
\affiliation{ITP, Universit\"{a}t Heidelberg, Philosophenweg 16, 69120 Heidelberg, Germany}
\email{maturi@uni-heidelberg.de}

\author[0000-0002-9489-7765]{Henry Joy McCracken}
\affiliation{Institut d’Astrophysique de Paris, UMR 7095, CNRS, and Sorbonne Université, 98 bis boulevard Arago, F-75014 Paris, France}
\email{hjmcc@iap.fr}

\author[0000-0002-3473-6716]{Lauro Moscardini}
\affiliation{University of Bologna - Department of Physics and Astronomy “Augusto Righi” (DIFA), Via Gobetti 93/2, I-40129 Bologna, Italy}
\affiliation{INAF- Osservatorio di Astrofisica e Scienza dello Spazio, Via Gobetti 93/3, I-40129, Bologna, Italy}
\affiliation{INFN- Sezione di Bologna, Viale Berti Pichat 6/2, I-40127 Bologna, Italy}
\email{lauro.moscardini@unibo.it}

\author[0000-0003-2397-0360]{Louise Paquereau} 
\affiliation{Institut d’Astrophysique de Paris, UMR 7095, CNRS, and Sorbonne Université, 98 bis boulevard Arago, F-75014 Paris, France}
\email{louise.paquereau29@gmail.com}

\author[0000-0002-4485-8549]{Jason Rhodes}
\affiliation{Jet Propulsion Laboratory, California Institute of Technology, 4800 Oak Grove Drive, Pasadena, CA 91001, USA}
\email{jason.d.rhodes@jpl.nasa.gov}

\author[0000-0003-0427-8387]{R. Michael Rich}
\affiliation{Department of Physics and Astronomy, UCLA, PAB 430 Portola Plaza, Box 951547, Los Angeles, CA 90095-1547, USA}
\email{rmrastro@gmail.com}

\author[0000-0002-4271-0364]{Brant E. Robertson}
\affiliation{Department of Astronomy and Astrophysics, University of California, Santa Cruz, 1156 High Street, Santa Cruz, CA 95064, USA}
\email{brant@ucsc.edu}

\author[0009-0007-4472-6136]{Samaneh Shamyati}
\affiliation{Department of Physics and Astronomy, University of California, Riverside, 900 University Ave, Riverside, CA 92521, USA}
\email{ssham033@ucr.edu}

\author[0000-0002-7087-0701]{Marko Shuntov}
\affiliation{Cosmic Dawn Center (DAWN), Denmark} 
\affiliation{Niels Bohr Institute, University of Copenhagen, Jagtvej 128, DK-2200, Copenhagen, Denmark}
\affiliation{University of Geneva, 24 rue du Général-Dufour, 1211 Genève 4, Switzerland}
\email{marko.shuntov@nbi.ku.dk}

\author[0000-0002-8437-6659]{Can Xu}
\affiliation{School of Astronomy and Space Science, Nanjing University, Nanjing, Jiangsu 210093, China}
\affiliation{Key Laboratory of Modern Astronomy and Astrophysics, Nanjing University, Ministry of Education, Nanjing 210093, China}
\affiliation{Kavli Institute for the Physics and Mathematics of the Universe (WPI), The University of Tokyo, Kashiwa, Chiba 277-8583, Japan}
\email{canxu@smail.nju.edu.cn}

\begin{abstract}

High-redshift ($z > 3$), massive quiescent galaxies (QGs) offer a significant window into early Universe galaxy formation. Previous works have predicted miscellaneous properties for these quiescents, from an overdensity of neighbors to elevated quenched fractions among such neighbors (i.e. galactic conformity). However, due to a scarcity in highly-resolved deep-field observations until recently, these properties have not been closely examined and pose unresolved questions for galaxy evolution. With new photometric-redshift catalogs from JWST data in the COSMOS-Web field, we present the S$\mathrm{\hat{O}}$NG sample, comprising 171 photometrically selected massive ($\geq10^{10}$ M$_\odot$) QGs with $3\leq$ $z\mathrm{_{phot}}$ $<$ 5. We look for low-mass neighbors around our sample and find substantial populations of star-forming galaxies (SFGs), contrasting the conformity effect at low-$z$. Our QGs also exhibit diverse clustering, from having no neighbors to potentially residing in environments no denser than star-forming equivalents, to being accompanied by SFGs with more stellar mass than the QG itself. Using a geometric method, we also report filamentary signals for 4\% of our sample, suggestive of some QGs' rejuvenation via cold gas accretion. We reapply the analysis on seven spectroscopically confirmed QGs in COSMOS-Web (M$_*$ $\sim$ $10^9-10^{11}$ M$_\odot$) and note similar patterns. Lastly, we report on Saigon, the most distant low-mass quiescent galaxy known to date ($z =$ 4.55, M$_*$ $= 1.33 \times10^9$ M$_\odot$); this spectroscopically confirmed QG resides in a protocluster candidate with 11 SFGs. These results pave new paths towards understanding QG environment, while also signaling an opportune era to examine their evolution with JWST.

\end{abstract}

\section{INTRODUCTION} \label{sec:intro}

In recent years, a number of deep near-infrared (NIR) surveys of the early Universe have found an abundance of massive galaxies (M$_*$ $\gtrsim$ 10$^{10}$ M$_\odot$) that suppressed most of their star formation---i.e. became quenched---as early as $z = 3-4$ \citep[e.g.,][]{Schreiber, Merlin2019, Valentino2020, AntwiDanso2025, Long2024, Carnall2024}. With the advent of the JWST Observatory, massive quiescent candidates were even discovered at $z\sim5-7$ \citep{Weibel2024, deGraaff2025}. For such extreme galaxies to exist, they would have had to undergo intense star formation, with star formation rates (SFRs)\,$>10^{2-3}$\,M$_\odot$\,yr$^{-1}$ over a short burst of $\sim$ 50 Myr \citep{Glazebrook2017, Valentino2020}, then followed by rapid quenching \citep[timescales $\sim100-300$ Myr;][]{Kakimoto2024, deGraaff2025}. Yet such phenomena are rarely (if at all) reproducible in current semi-analytical models and cosmological simulations, where galaxies can indeed reach similarly high stellar masses at early redshifts, but only quench at much later epochs \citep[e.g.,][]{Merlin2019, Forrest2020a, Lovell2023}. These discrepancies signal potential issues with our current models and call for further inquiry into the physical processes driving massive quiescent galaxy (QG) evolution at high-$z$ \citep{Lagos2025}.

A widely discussed source of influence on the evolution of massive quiescent galaxies is the local environment. Major mergers, which occur between galaxies similar in M$_*$ (mass ratio $\geq$ 1$:$4 between a `host' and its `neighbor'), can change the morphology of star-forming disks and transform them into compact, bulge-dominated quiescents \citep[e.g.,][]{Hopkins2006}. While such drastic changes are instrumental to the initial creation of massive QGs, their long-term impact pales in comparison to the more frequent minor mergers between galaxies of significant mass discrepancies (mass ratio $<$ 1$:$10). Firstly, ``dry'' (i.e. gas-poor) minor mergers are hypothesized to make massive QGs increase in size, although studies disagree on whether such mergers are single-handedly responsible for the rapid size evolution observed at high-$z$ or require additional physical processes \citep[e.g.,][]{Bezanson2009, Newman2012}. Secondly, dry minor mergers can introduce new stellar populations into QGs that will alter their overall color gradients over time \citep{Suess2020, Suess2023}. Lastly, in limited cases, ``wet'' (i.e. gas-rich) minor mergers can reignite quiescent galaxy star formation, in a process known as rejuvenation \citep{Kaviraj2014}. In summary, the impact of minor mergers on quiescents can be significant in various ways, despite the inferior masses of the merging neighbors. The reason is but a matter of statistics: according to the $\Lambda$CDM model, within a dark matter (DM) halo, there are a greater number of low-mass DM sub-halos than high-mass ones \citep{PressSchechter, Jiang2014}, suggesting that galaxies in dense groups or clusters are more likely to have large mass discrepancies with their host galaxies, i.e. minor mergers have a higher likelihood of occurring \citep{Fakhouri2010}. And since QGs at $z$ $\sim$ 1 are predominantly detected in those same dense environments \citep[e.g.,][]{Kauffmann2004, Peng2010}, looking for signs of minor mergers at high-$z$ can offer insights into QG evolution and thus improve our overall models of massive galaxy assembly.

Besides influencing the evolution of the central galaxy, interactions between a QG and its neighbors can correlate with effects on the neighbors themselves. The most researched effect at low-$z$ is ``galactic conformity'' \citep{Weinmann2006, Knobel2015, Ayromlou2023}, whereby neighbors around a massive QG have a higher likelihood of becoming quenched than neighbors around a star-forming galaxy (SFG) at similar redshifts. It is not yet apparent as to what motivates this conformity; explanations offered for this effect have ranged from the interstellar medium simply being ``pre-heated'' at early epochs \citep{Kauffmann2013} to satellite galaxies experiencing ram-pressure stripping as they move through such warm interstellar gas \citep{Ayromlou2023}. Regardless, conformity is observed in galaxy groups up to $z\sim 2$ \citep{Hartley2015, Kawinwanichakij2016, Treyer2018}, but is not well-established past this epoch. This raises the question of when conformity sets in over the lifetime of a central galaxy. \cite{Hearin2016} predict that conformity signals will weaken significantly beyond $z$ $\sim$ 1, due to correlations between the specific star formation rate (sSFR $=$ SFR/$\mathrm{M_*}$) of a central galaxy and its halo accretion rate. However, high-$z$ QGs themselves are not only uncommon, but also historically difficult to confirm. QG samples compiled in the era before JWST are often contaminated by dusty SFGs, the fraction of which can be up to $\sim$15\% at $z$ $\gtrsim$ 2 \citep{Man2016}. One major reason is that the selection boundary which separates QGs from SFGs on color-color diagrams is empirically derived \citep{Williams2009, Ilbert2013}, which means the threshold may not always hold in the face of new data or population diversity; and through that leeway, dusty SFGs have sometimes intruded into quiescent space \citep{Hwang2021}. In general, until very recently, even selecting robust samples of high-$z$ QGs has been an arduous task, let alone measuring the star formation rates of their neighbors and assessing conformity.

But discussions of mergers and conformity also hinge on a critical question: just how many neighbors should we expect around high-$z$ QGs in the first place? A common way to quantify this abundance has been through the clustering signal, i.e., how concentrated the neighbors are around their central QG. This clustering signal is closely linked to the halo occupation distribution (HOD)---the distribution of galaxies within dark matter halos based on an assumed connection between halo clustering and the physics of galaxy formation \citep[e.g.,][]{BerlindandWeinberg2002, Kravtsov2004, Zheng2005}. At $z$ $<$ 3, massive QGs are found to have stronger clustering signals in their neighborhood (i.e., have more neighbors) than stellar-mass-matched SFGs \citep[e.g.,][]{Kawinwanichakij2014, Berti2019}, and simultaneously, QGs tend to live within more massive halos than SFGs \citep[e.g.,][]{Velander2014, Mandelbaum2016}. When considered together, these two phenomena make sense physically: with their halos being more massive than those of their star-forming counterparts, low-$z$ QGs will gravitationally attract more objects towards themselves. But is this also true at high-$z$, where both star formation and quenching occurred very rapidly for the first QGs in the Universe? A second question also arises: is it the halo mass or the evolutionary stage of a massive QG that governs its clustering signal?

As it stands, the discourse on massive QGs and low-mass neighbors in relation to one another remains underdeveloped at $z>2$, when most of these QGs are believed to form \citep[e.g.,][]{Thomas2005, Citro2016}. The most critical reason for this lies with the lack of sensitivity pre-JWST. Even extensive deep-field surveys like CANDELS \citep{CANDELS2011}, which had some of the deepest HST imaging, could only investigate systems with mass ratios $\gtrsim$ 1:10 and redshifts $\leq$ 2 \citep{Newman2012}. Despite isolated successes from ground-based surveys \citep[e.g.,][]{Ito2023, Tanaka2024}, the overall poor sensitivity has fogged our view of the red and/or low-mass neighbors predicted to be profuse in the early Universe. Thus, the high-$z$ regime ($z \sim 2-$3), where merger rates are theorized to peak \citep{Wetzel2009}, remains weakly constrained in observations.

But whatever metaphorical fog has been hanging over our sight, JWST has certainly started to clear it.\footnote{The phrasing here echoes a couplet on opportune new beginnings from Nguyễn Du's classic \textit{The Tale of Kiều}: ``Trời còn để có hôm nay/Tan sương đầu ngõ, vén mây giữa trời'' (Heaven has let a day like this come by/Clears fog in alleys, tucks clouds in the sky). The English translation is from \cite{Kieu}.} Since 2021, thanks to the telescope's enhanced sensitivity, IR wavelength coverage and high spatial resolution, we are now unraveling substantial populations of massive QGs at $z>3$ in overdensities \citep[e.g.,][]{deGraaff2025, McConachie2025}, some of which have very low mass ratios of $\lesssim$ 1:100, well below what was previously achieved by HST \citep{Suess2023}. Additionally, with the success of extensive surveys like COSMOS-Web \citep{COSMOSWeb2023} and CEERS \citep{CEERS}, we can carry out large statistical surveys of both QGs and SFGs to quantify how the local environment affects galaxy growth over cosmic time.

Given these new exciting opportunities, we are seeking to address long-standing questions on the environments of massive QGs in the early Universe, using a photometrically-derived sample of $n = 171$ massive QGs at $z=3-5$. In $\S$\ref{sec:data}, we describe the COSMOS-Web observational data that we use for our analysis. In $\S$\ref{sec:search}, we enumerate the steps taken to compile a robust catalog of neighbors. In $\S$\ref{sec:results}, we examine these neighbors in terms of physical properties, clustering signals, and signs of large-scale structures. We also discuss broader implications for early QG evolution, including the emergence of protoclusters, the rejuvenation of star formation, and potential quenching pathways. In $\S$\ref{sec:specz}, we apply our method to a sample of spectroscopically-confirmed massive QGs at $z\gtrsim 3$, and summarize our findings in $\S$\ref{sec:sumcon}. Throughout our work, we assume a concordance $\Lambda$CDM cosmology with $\Omega_m = 0.3$, $\Omega_{\Lambda} = 0.7$, and $H_0$ $=$ 70 km s$^{-1}$ Mpc$^{-1}$.

\section{DATA \& PHOTOMETRIC SAMPLE} \label{sec:data}
  
\subsection{COSMOS-Web} \label{COSMOS-Web}

COSMOS-Web is a wide-field JWST Cycle 1 treasury program that builds upon the rich, multiwavelength archival observations in the COSMOS field \citep{2007Scoville}. It is the largest contiguous area survey with JWST thus far, mapping 0.54 deg$^2$ using four NIRCam filters (F115W, F150W, F277W, and F444W) to 3$\sigma$ depths of $\sim$ 26.9$-$28.8 AB, and 0.19 deg$^2$ using one MIRI filter (F770W) to a 3$\sigma$ depth of  $\sim$ 25.5 AB\citep{COSMOSWeb2023}. The data reduction process for NIRCam and MIRI has been fully detailed in \cite{Franco2025} and \cite{Harish2025}, respectively, but will also be briefly described here. The NIRCam images were reduced using version 1.12.1 of the JWST Calibration Pipeline, with additional customizations following other JWST NIRCam surveys \citep[e.g.,][]{Bagley2022}. Using the Calibration Reference Data System (CRDS) pmap 1170, the final mosaics were generated with a pixel scale of $0 \farcs 03$ in Stage 3 of the pipeline. Lastly, we perform astrometric calibrations with the JWST/HST alignment tool, based on an HST/F814W $0\farcs03$-scale mosaic in the COSMOS field \citep{Koekemoer2007}, which is calibrated to Gaia-EDR3 \citep{Gaia2018}. In comparison to the NIRCam mosaic, our reference catalog differs by a median offset of $< 5$ mas. The MIRI images were reduced using version 1.12.5 of the JWST Calibration Pipeline, with additional background subtraction to reduce instrumental noise. The MIRI mosaic was resampled onto a common output grid with a pixel size of $0\farcs03$/pixel and aligned with HST/F814W images of the same region.

Our work relies on the COSMOS2025 photometric catalogs\footnote{https://cosmos2025.iap.fr/} from COSMOS-Web, which are described in detail in \cite{Shuntov2025}. In summary, we use the $\texttt{SourceXtractor++}$ package \citep[][]{1996A&AS..117..393B, 2016JOSS....1...58B} to perform source detection on a $\chi^2$ detection image obtained by co-adding all the available NIRCam bands. Using a ``hot-and-cold'' detection scheme, photometry is measured over fixed apertures, as well as 2D models convolved with a generated PSF. Then, using the same package, we perform simultaneous and iterative fitting on neighboring sources, which are grouped together to ensure more accurate, less contaminated fluxes and shapes. Lastly, we calibrate model-based photometric uncertainties to account for both source photon noise and background Poisson noise. We also clean the catalog and flag any sources that, among others, are `hot pixels', have significant inconsistencies between space- and ground-based band photometry, or show other signs of problematic photometry and derived $z\mathrm{_{phot}}$ (e.g., contaminant diffraction spikes from stars).

At the redshift range relevant to our work ($z\mathrm{_{phot}}=3-6$), the stellar mass completeness limit of COSMOS2025 is $\mathrm{M_*} \sim 10^{8.1}$ $\mathrm{M_\odot}$ for $z$ $=$ 3, and $\mathrm{M_*} \sim 10^{8.6}$ $\mathrm{M_\odot}$ for $z$ $=$ 6. This makes COSMOS2025 a well-suited catalog for large-scale statistical studies of early-Universe galaxy populations, including minor mergers with mass ratios $<$1$:$10.
  
\subsection{Quenched galaxy sample construction} \label{subsec:sample-construction}

Before selecting QGs, we preemptively crosscheck for detections in the (blind extraction) A$^3$COSMOS catalog \citep[version 20220606,][]{Adscheid2024}, a compilation of archival Atacama Large Millimeter Array (ALMA) observations over the COSMOS field. The purpose is to discern potential dusty submillimeter galaxies that may have been erroneously fit as high-$z$ QGs. This yields no matches. We also crosscheck for active galactic nuclei (AGN) contamination by comparing our source list to Chandra X-ray data, as well as radio data from MIGHTEE \citep{MIGHTEE}, VLA 1.4 GHz \citep{VLA2007}, and VLA 3 GHz \citep{VLA2017}---all of which are surveys done on the COSMOS field. For all cross-catalog checks throughout this work, our search radius is 1$''$.

Finally, we turn back to COSMOS2025 and filter the catalog for sources that meet the requirements for robust redshifts and clean photometry. We select only sources whose \texttt{warn\_flag} is marked 0 \citep[i.e. sources that are considered `most secure' by][]{Shuntov2025} and whose AB magnitude in the F444W band is $<$\,30 in the photometry catalog.

\begin{table*}[t]
\centering 
\hspace*{-1.5cm} 
\begin{tabular}{llllllr}
\toprule
Name & R.A. & Decl. & $z_\mathrm{phot}$ & log$_{10}$(M$_*$/M$_\odot$) & SFR(M$_*$/yr) & log$_{10}$(sSFR/yr) \\
\midrule
SÔNG-1 & 09:59:04.64 & $+$02:09:06.45 & $3.26_{-0.09}^{+0.1}$ & 11.6 $\pm$ 0.1 & 11 $\pm$ 7 & $-$10.5 \\
SÔNG-2 & 09:59:03.45 & $+$02:10:25.43 & $3.05_{-0.04}^{+0.04}$ & 10.7 $\pm$ 0.1 & $<$ 1 & $-$11.2 \\
SÔNG-3 & 09:59:21.46 & $+$02:09:42.23 & $3.2_{-0.04}^{+0.05}$ & 10.6 $\pm$ 0.1 & $<$ 1 & $-$12.4 \\
SÔNG-4 & 09:59:28.86 & $+$02:11:07.87 & $3.39_{-0.06}^{+0.18}$ & 11.1 $\pm$ 0.1 & 12 $\pm$ 5 & $-$10.0 \\
SÔNG-5 & 09:59:03.27 & $+$02:14:14.76 & $3.35_{-0.04}^{+0.04}$ & 11.1 $\pm$ 0.1 & 1 $\pm$ 1 & $-$11.0 \\
SÔNG-6 & 09:59:21.08 & $+$02:13:45.12 & $3.17_{-0.1}^{+0.05}$ & 10.6 $\pm$ 0.1 & $<$ 1 & $-$12.9 \\
SÔNG-7 & 09:59:34.41 & $+$02:13:18.23 & $3.48_{-0.13}^{+0.13}$ & 10.6 $\pm$ 0.1 & $<$ 1 & $-$11.4 \\
SÔNG-8 & 09:59:32.62 & $+$02:16:33.72 & $3.3_{-0.02}^{+0.02}$ & 10.6 $\pm$ 0.1 & 2 $\pm$ 1 & $-$10.2 \\
SÔNG-9 & 09:59:53.89 & $+$02:04:17.05 & $3.08_{-0.14}^{+0.12}$ & 10.5 $\pm$ 0.1 & 1 $\pm$ 1 & $-$10.5 \\
SÔNG-10 & 09:59:36.6 & $+$02:05:59.38 & $3.28_{-0.09}^{+0.37}$ & 11.2 $\pm$ 0.1 & $<$ 1 & $-$11.4 \\ 
\bottomrule
\end{tabular}
\caption{\label{Table:galaxysample} A snippet of the S$\mathrm{\hat{O}}$NG sample. The photometric redshift ($z_\mathrm{phot}$) is the \texttt{LePHARE} SED-fitting output from \cite{Shuntov2025}, and is reported with both the $-\sigma$ (subscript) and $+\sigma$ (superscript) values. The stellar mass ($\mathrm{M_*}$) and star formation rate (SFR) are \texttt{CIGALE} outputs from \cite{ArangoToro2025}. The specific star formation rate (sSFR) is from this work. The full data has been released along with this paper.}\
\end{table*}
    
Now, we identify massive QGs in COSMOS-Web. We first search for sources that have $\mathrm{M_*}$ $\geq$ 10$^{10}$ M$_\odot$ in \textit{both} galaxy catalogs generated by \cite{Shuntov2025} and \cite{ArangoToro2025}, which use $\texttt{LePHARE}$ \citep{Arnouts2002, Ilbert2006} and $\texttt{CIGALE}$ \citep{CIGALE}, respectively. Then, we set our quenching threshold by the specific star formation rate such that $\mathrm{log_{10}(sSFR)}$ $\leq -9.8$ yr$^{-1}$ \citep[see e.g.,][]{Bisigello2018, CarnallQG} and find sources whose sSFR satisfies this criterion in \textit{both} the $\texttt{LePHARE}$ and $\texttt{CIGALE}$ catalogs. For our analysis, we consider only sources with $\texttt{LePHARE}$ photometric redshifts at $3 \leq z_\mathrm{phot}<7$. Finally, we utilize $\texttt{CIGALE}$'s $\chi^2$ values to devise another criterion for robustness, selecting only sources with reduced $\chi^2 \leq 10$. Note that we select from both $\texttt{LePHARE}$ and $\texttt{CIGALE}$ outputs because $\texttt{LePHARE}$ mainly serves to estimate $z\mathrm{_{phot}}$, while $\texttt{CIGALE}$ uses $\texttt{LePHARE}$'s $z\mathrm{_{phot}}$ but also incorporates non-parametric star formation histories (SFH) and different attenuation laws, allowing more precise estimates of physical properties such as M$_*$ and SFR \citep{ArangoToro2025}. This reasoning holds for both this QG identification and the neighbor search in $\S$\ref{sec:search}.

After our filtering process, we find 172 QGs. We cross-match our sources to the COSMOS-Web spectroscopic redshift catalog by \cite{Khostovan2025} and only identify two with reliable $z\mathrm{_{spec}}$ ($Q_f$ $=$ 2$-$4). One such system, with $z\mathrm{_{spec}}$ $=$ 3.456 from the MOSDEF survey \citep{MOSDEF}, is surprisingly detected by the COSMOS-Web segmentation map as two distinct galaxies at different redshifts ($z\mathrm{_{phot}}$ $=$ 1.36 and $z\mathrm{_{phot}}$ $=$ 3.29). Given the system's highly disturbed morphology and the clear tidal bridges between the segments that COSMOS-Web identifies as two separate galaxies\footnote{In COSMOS2025, the segment identified as a low-$z$ galaxy has ID 667931, and the supposed high-$z$ QG has ID 667932.}, we strongly suspect that this is instead a dusty major merger system masquerading as a high-$z$ QG. Thus, we remove the object from our analysis, and note that its removal or inclusion does not affect our final results thanks to our large sample size. This leaves us with 171 QGs, only one of which has a reliable $z\mathrm{_{spec}}$ from MAGAZ3NE \citep{Forrest2020b}. In the forthcoming neighbor search, we will analyze this one source using its $z\mathrm{_{spec}}$.

We continue with 171 QGs, which we call the S$\mathrm{\hat{O}}$NG sample, after the Vietnamese word for rivers.\footnote{Inspired by \textit{thiên hà} (sky river), an epithet of the Milky Way attested in Vietnamese literature since at least the fourteenth century, later adopted as the astronomical term for galaxies in Hoàng Xuân Hãn's \textit{Danh từ khoa học: Vocabulaire scientifique} (1942).} In general, these galaxies have $\mathrm{M_*}$ $=$ $10^{10}-10^{11.6}$ ~$\mathrm{M_\odot}$, with a median of $4.13_{-1.89}^{+4.93} \times10^{10} \mathrm{M_\odot}$; 35.5\% have $\mathrm{M_*}$ $\leq$ $10^{10.5}$ $\mathrm{M_\odot}$. Their sSFR median is log$_{10}$(sSFR/yr) $= -11.1_{-0.9}^{+0.8}$. In terms of redshift, they have $z\mathrm{_{phot}}$ $=$ $3.00-4.58$, with a median of $z\mathrm{_{phot}} = 3.35_{-0.24}^{+0.55}$ and a median uncertainty of $\sigma_z$ $=$ 0.079, much more restricted than the wide net which we initially cast ($3 \leq z_\mathrm{phot}<7$). For an extensive catalog of these QGs' basic properties, see Table \ref{Table:galaxysample}. Before we proceed, it is worth noting that of our QGs with \texttt{LePHARE} $z\mathrm{_{phot}}$ $\leq$ 3.7, a staggering 92\% are identified as potential members in the COSMOS-Web group and protocluster core catalog, while 17\% are \textit{robust} members, with associated probabilities $\geq$ 0.5 \citep{Toni2025}. This has promising implications for our search.

\section{NEIGHBOR SEARCH} \label{sec:search}

In this part, we narrow down step-by-step the sources that have potential physical associations to the S$\mathrm{\hat{O}}$NG QGs. Our basis is the photometry-only COSMOS2025 catalog, which, thanks to NIRCam, has shown a factor-of-two improvement over COSMOS2020 in terms of observational depth at 26 AB mag \citep[see][]{Shuntov2025}. Since our high-$z$ neighbors are likely to be small, faint, even blended with their brighter hosts, we expect that some of the neighbors may be resolved by NIRCam, but not by ground-based telescopes, which leads to large discrepancies between ground- and space-based observations that will cause these neighbors to be flagged as `unreliable' in the main galaxy catalog. By instead considering the photometry-only COSMOS2025 catalog, we hope to circumvent this potential incompleteness.

Our selection process relies on multiple criteria: (1) the angular distance between each source and the central galaxy; (2) the number of significant UV-to-NIR detections for each source; and (3) the redshift difference between each source and the corresponding central galaxy, using photometric redshifts ($z_\mathrm{phot}$) from \texttt{LePHARE}.

\subsection{Angular distance to central galaxy} \label{angular}

In the first step of our procedure, we search through the COSMOS2025 photometry catalog for all NIRCam sources within a specified radius of each of our 171 QGs' coordinates. The radius that we adopt is derived from the virial radius, r$_\mathrm{vir}$ for M$_\mathrm{halo}$ $\sim$ 10$^{13}$ M$_\odot$, which is the typical halo mass estimate for QGs with stellar masses M$_*$ $\sim$ 10$^{11}$ M$_\odot$ at $z$ $>$ 2 \citep{Behroozi2010}. There are still uncertainties as to whether massive halos or galaxy protoclusters at these epochs have been confined inside a virial radius or are still extending beyond it \citep[see e.g.,][]{Chiang2013, Chiang2017}. However, to be conservative with our search, we adopt a radius that can identify the closest possible associated sources for our QGs. We set that radius to be 25\% of our calculated r$_\mathrm{vir}$, which translates to a comoving distance of $\sim$ 100\,kpc, or 15$''$ at $z\sim3.5$. Applying this 15$''$ radius around each QG, we obtain an initial catalog of $\sim$14,400 neighbor candidates.

\subsection{Number of significant UV-to-NIR detections} \label{threedetections}

Expecting faint, low-mass neighbors, we set a second criterion: in a given band, a detection is significant if SNR $>3$. Then, to guarantee real sources, we only keep those with three or more detections across these bands: HST ACS (F814W), JWST NIRCam (F115W, F150W, F277W, F444W), JWST MIRI (F770W), CFHT MegaCam (u), Subaru HSC (\textit{g}, \textit{r}, \textit{i}, \textit{z}, \textit{y}), VISTA VIRCAM (Y, J, H, Ks), and Spitzer IRAC (CH1, CH2, CH3, CH4). After applying this check to the results of $\S$\ref{angular}, the number of candidates remains the same at $\sim$14,400.

\subsection{Photometric redshifts \& SED fitting} \label{subsec:zSED}

For this criterion, we use spectral energy distribution (SED) fits from \cite{Shuntov2025}, which are generated by \texttt{LePHARE} \citep{Arnouts2002, Ilbert2006}. $\texttt{LePHARE}$ works by comparing predicted magnitudes and/or fluxes for various types of galaxies and AGN templates with the observed photometric fluxes that are provided as inputs.

Many of the specifics on the \texttt{LePHARE} parameters used for COSMOS2025 have been expounded in \cite{Shuntov2025}; however, we will also provide an overview in relation to our own analysis. We adopt the \cite{2003MNRAS.344.1000B} stellar population synthesis models, along with templates used by \cite{Ilbert2015}, which in tandem allow us to compute physical parameters simultaneously on the basis of a more diverse template library. Our models are fit over a wide range of ages (0.05$-$13.5 Gyr) and two metallicities (solar and half-solar). Additionally, we account for dust by including three different attenuation curves \citep{2000ApJ...533..682C, 2013A&A...558A..67A, Salim2018}, as well as factoring in the effects of polycyclic aromatic hydrocarbon (PAH) emission. This incorporation of dust proves to be necessary: there are galaxies suggested to be high-$z$ massive QGs in past spectroscopic studies that have since been re-identified as dusty low-$z$ interlopers \citep[e.g.,][]{InterloperPaper}.

Before we proceed with our analysis using the \texttt{LePHARE} SED fits, we also recognize that some of these sources ($n=$ 236) have their \texttt{warn\_flag} column marked as either 2 or 3 in COSMOS2025, which indicates major discrepancies between their ground- and space-based detections. As we have discussed at the beginning of $\S$\ref{sec:search}, this might be because some of our sources are faint and have not been well-resolved in ground-based observations. To account for any imprecise fitting that the discrepancies may have caused, we isolate the space-based fluxes of the discrepant sources (both from HST and JWST), then rerun \texttt{LePHARE} via the GAZPAR\footnote{https://gazpar.lam.fr/home} interface with the same setup as \cite{Shuntov2025}. These new outputs are considered in lieu of the \cite{Shuntov2025} results only for the discrepant sources. Last but not least, in the hopes of finding robust neighbors, we simultaneously crosscheck our candidates from $\S$\ref{threedetections} with the \cite{Khostovan2025} $z\mathrm{_{spec}}$ catalog and will consider any reliable $z\mathrm{_{spec}}$ ($Q_f$ $=$ 2$-$4) in lieu of the $z\mathrm{_{phot}}$ if the opportunity arises.

\begin{figure*}[p] 
    \centering
      \subfloat[a][]{\includegraphics[scale=0.724]{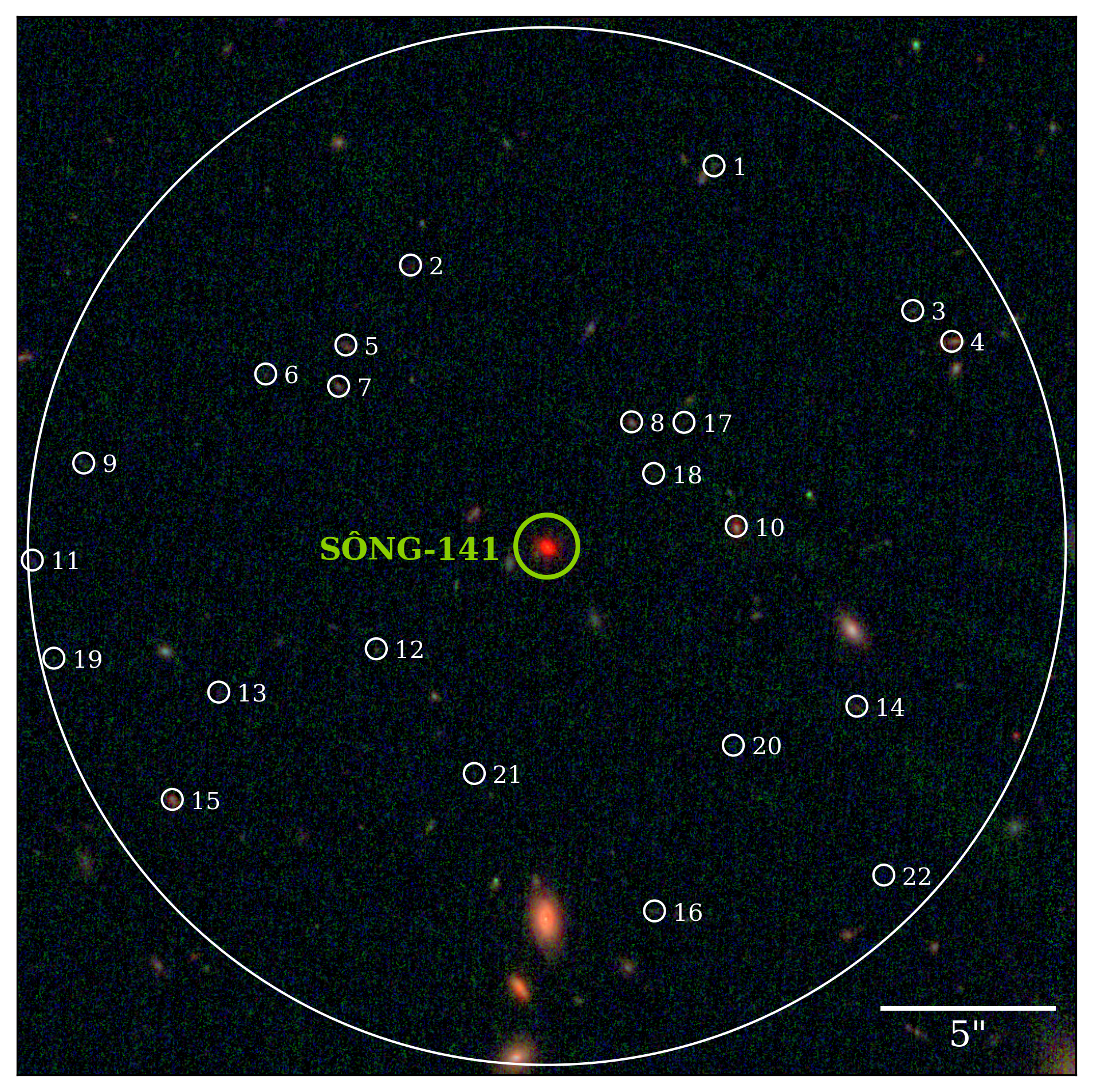} \label{fig:song-a}} \\
      \subfloat[b][]{\includegraphics[scale=0.32]{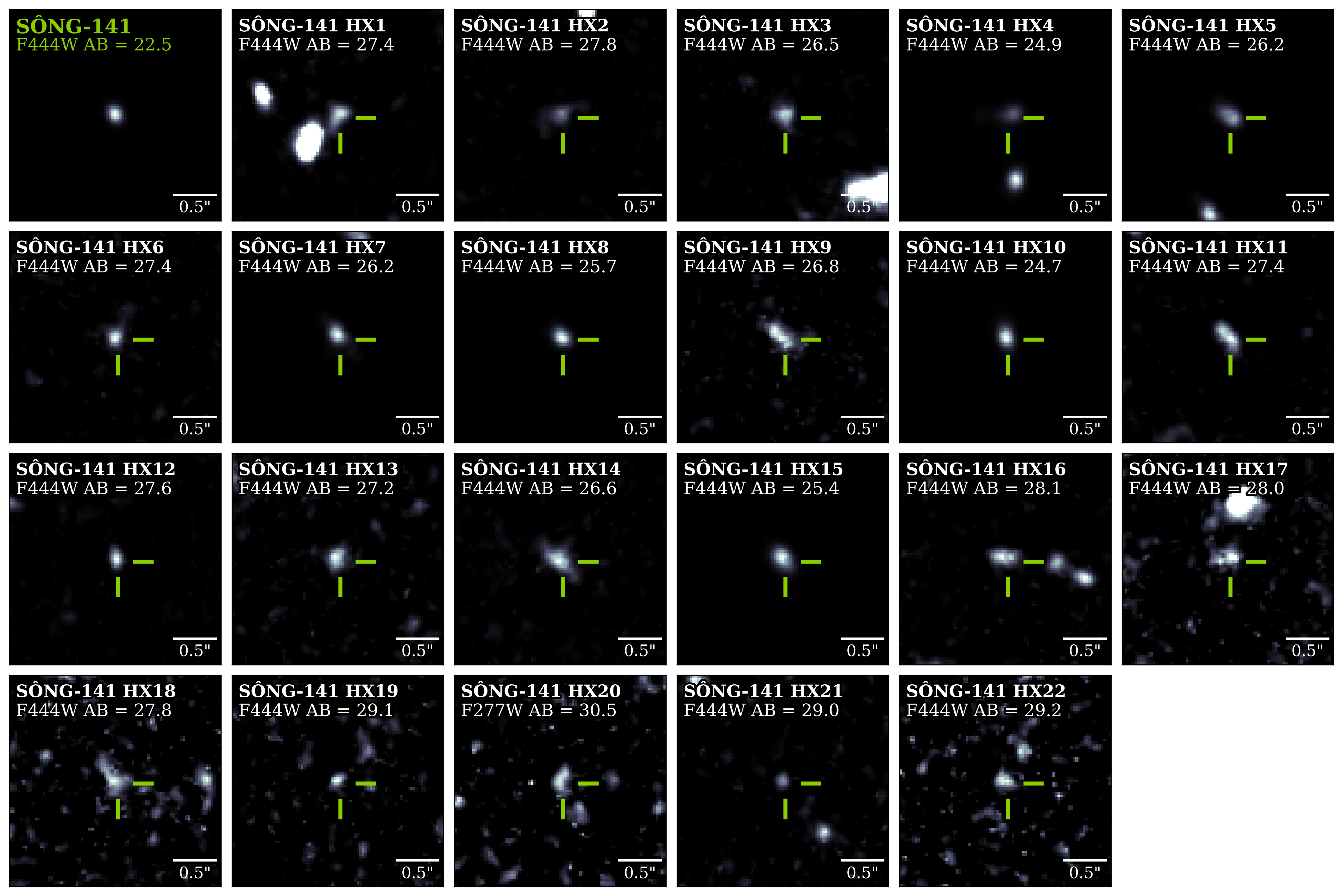} \label{fig:song-b}}
      \caption{(a) RGB image of S$\mathrm{\hat{O}}$NG-141, one QG at $z =$ 3.63 (circled green at the center). The bands chosen for R, G, and B are NIRCam F444W, F277W, and F150W, respectively. All neighbors within 15$''$ are circled white, their names abbreviated as numbers on their right. (b) $\chi^2$ detection images of S$\mathrm{\hat{O}}$NG-141 and its neighbors, constructed and reduced from all four NIRCam bands (F115W, F150W, F277W, and F444W). The image contrast has been manually adjusted for visibility. Each galaxy's AB magnitude in the reddest band available is annotated on the top left corner of its corresponding cutout.} \label{fig:SONG-154_2figs}
\end{figure*}

With the \texttt{LePHARE} outputs from both COSMOS2025 and GAZPAR, we define a new criterion based on the $1\sigma$ uncertainty in $z_\mathrm{phot}$\textemdash a method also used in other neighbor searches such as \cite{Suess2023}. Our criterion is as follows: the $z_\mathrm{phot}$ of both the host and each of its neighbors must be within $1\sigma$ of one another, such that $z_\mathrm{QG}$ $\pm$ $\sigma_\mathrm{QG}$ must overlap with $z_\mathrm{neighbor}$ $\pm$ $\sigma_\mathrm{neighbor}$. We use the asymmetric $z_\mathrm{phot}$ uncertainties for the upper and lower 16$^\mathrm{th}$ and 84$^\mathrm{th}$ percentiles for each source. If a host or a neighbor has reliable $z\mathrm{_{spec}}$ in \cite{Khostovan2025}, we check to see if the $z\mathrm{_{spec}}$ of that host/neighbor is within 1$\sigma$ of the $z_\mathrm{phot}$ of its neighbors/host. If a host \textit{and} a neighbor both have $z\mathrm{_{spec}}$, we single them out for inspection. In the end, of the $\sim$14,400 sources identified in $\S$\ref{threedetections}, only $\sim$3,300 pass this redshift criterion. 

Since brown dwarfs in our own galaxy can sometimes mimic faint red galaxies  \citep[a trend already noted in the recent literature; see][]{Burgasser2024, Hainline2024, Tu2025}, we also fit brown dwarf models to these sources to identify potential contaminants. We follow the procedure described in \cite{AkinsBrownDwarf}, then compare the brown dwarf fits to the previous galaxy fits in terms of $\chi^2$. A source is considered misidentified if its brown dwarf $\chi^2$ is smaller than its $\texttt{CIGALE}$ galaxy $\chi^2$ as reported by \cite{ArangoToro2025}. We then gather all the galaxies that pass this check and make one last robust cut, selecting only galaxies with reduced $\chi^2 \leq 10$, mirroring our procedure in $\S$\ref{subsec:sample-construction}. Lastly, we remove nine sources with $\texttt{CIGALE}$ $\mathrm{M_*}$ $<$ 1 $\mathrm{M_\odot}$. This peculiarity may stem from $\texttt{CIGALE}$'s inability to return physical $\mathrm{M_*}$ estimates and does not necessarily negate the integrity of a source; still, we add this check to ensure a conservative sample and facilitate the mass-based analysis in $\S$\ref{sec:results}. With these eliminations, we arrive at a catalog of 2,048 neighbors. Fig.~\ref{fig:SONG-154_2figs} is an RGB image of one QG and its identified neighbors; note the large outer circle in Fig.~\ref{fig:song-a}, which denotes the 15$''$ radius described in $\S$\ref{angular}.

\section{RESULTS \& DISCUSSION} \label{sec:results}

Around the 171 S$\mathrm{\hat{O}}$NG galaxies, we report a total of 2,048 neighbors. The general properties of our QG sample, both from photometric data and SED fits, have been made publicly available with this paper. We catalog the neighbors based on their host QG: each neighbor's name consists of its host's name, the suffix ``HX'', and a number. In the upcoming subsections, we discuss the resulting neighbor properties in detail. Section $\S$\ref{subsec:prop} summarizes trends in $\mathrm{M_*}$, morphology, and sSFR, and considers implications for conformity. Section $\S$\ref{subsec:clustering} examines clustering signals and compares our results against the background COSMOS field and a semi-analytical model. Section $\S$\ref{subsec:filament} presents a novel method to identify potential filamentary structures.

\begin{figure}
	\includegraphics[width=\columnwidth]{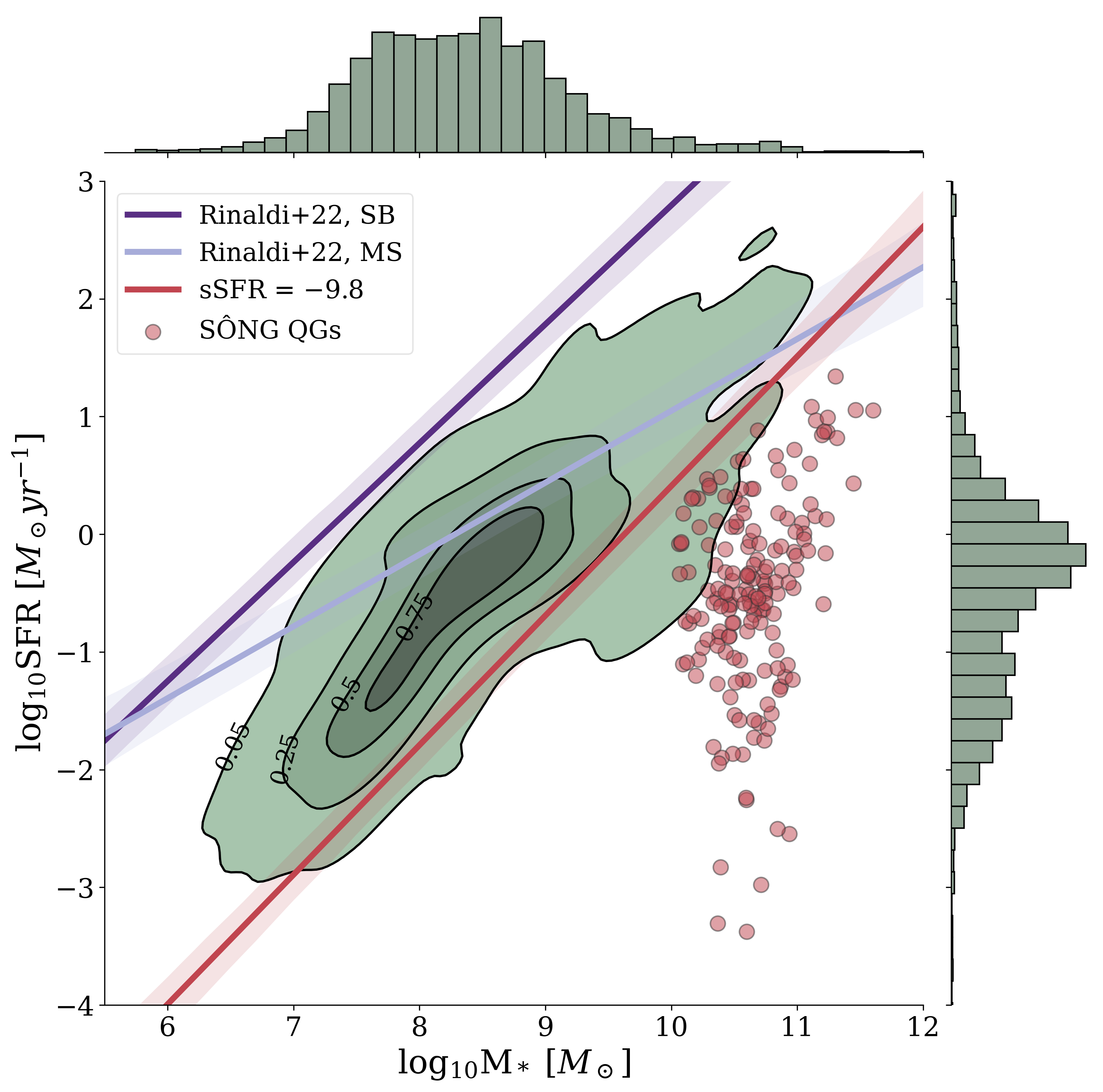}
    \caption{SFR vs. $\mathrm{M_*}$ for the S$\mathrm{\hat{O}}$NG QGs (red dots) and their neighbors (green density contours). The contour levels are iso-proportions, representing the fraction of neighbors outside each contour line; for example, a contour level of 0.05 means that 5\% of the neighbors are outside that line. The red straight line denotes our quenching threshold ($\mathrm{log_{10}(sSFR)}$ $=-9.8$ yr$^{-1}$); the purple and lavender straight lines denote the \cite{Rinaldi2022} parameterized best fits for starburst (SB) and main-sequence (MS) galaxies at $2.8 \leq z<4$. 
    For all the straight lines, 1$\sigma$ noise is added as shading. The top and right histograms show the $\mathrm{M_*}$ and SFR distributions.}
    \label{fig:mstar_vs_sfr}
\end{figure}

\subsection{General neighbor properties} \label{subsec:prop}

A brief census shows that of our S$\mathrm{\hat{O}}$NG sample, 97\% of QGs have more than one neighbor. 80\% of the QGs have the highest $\mathrm{M_*}$ within their respective search radius (i.e. they are likely to be the `centrals'), 16\% have one more massive neighbor, 3.5\% have two more massive neighbors. The neighbors have median $\mathrm{M_*}$ $=$ 2.19 $\times$ $10^8$ $\mathrm{M_\odot}$; only 4.5\% have mass ratios $>$ 1$:$4, the threshold for major mergers, while $\sim$63\% have mass ratios $<$ 1$:$100, far below the minimum threshold that HST is able to probe. A comprehensive illustration of the neighbor mass-SFR distribution is given in Fig.~\ref{fig:mstar_vs_sfr}. On a final note, only four of our neighbor sample have reliable $z\mathrm{_{spec}}$ in \cite{Khostovan2025}. We find no system where a host and a neighbor both have $z\mathrm{_{spec}}$.

We continue by cross-matching our results with the Adaptive Matched Identifier of Clustered Objects (AMICO) catalog from \cite{Toni2025}, a compilation of COSMOS-Web galaxy groups up to $z = 3.7$ constructed using a matched filtering algorithm. Of all our neighbors with \texttt{LePHARE} $z\mathrm{_{phot}}$ $\leq$ 3.7, 42\% were identified as potential members of $z$ $\leq$ 3.7 galaxy groups and protocluster cores in \cite{Toni2025}, though only 3\% make the cut for robust membership (associated probability $p$ $\geq$ 0.5). Still, our results expand upon \cite{Toni2025} and show that these galaxies are not just associated, but constitute larger protoclusters or protocluster cores wherein mass-dominant quiescent centrals are encircled by rich populations of low-mass neighbors. This indicates that the high-$z$ Universe already features well-organized group environments, consistent with hierarchical growth models.

After making RGB visualizations of the neighbors using NIRCam bands, we perform a quick, qualitative visual inspection. We find some visual traces of morphological structure. An illustrative example is in Fig.~\ref{fig:song-b}, where S$\mathrm{\hat{O}}$NG-141 HX11 is elongated like an edge-on disk galaxy. Most other neighbors are clumpy and irregular, with some showing what looks like ongoing mergers (e.g. S$\mathrm{\hat{O}}$NG-141 HX18, which is elongated and has what looks like a possible infalling companion). There is rich literature showing that mergers can induce starbursts, which, in tandem with AGN feedback, cause a rapid exhaustion of molecular gas \citep[e.g.,][]{Bower2006, Croton2006, Dave2017, Davies2022}. Hence, these possible signs of mergers may also be predictors that some neighbors will become quenched at lower $z$. Still, we must note that none of the neighbors are well-resolved enough for precise visual inspections. Further deep-field observations and more informed analytical studies are needed to draw firm conclusions on their morphology and its implications.

Based on \texttt{CIGALE} outputs, we find the neighbors' median SFR to be $4.13 \times 10^{-1}$ M$_\odot$ \,yr$^{-1}$. We do not discern any radially increasing or decreasing trends in the neighbors' SFR or $\mathrm{M_*}$ with respect to their distance from the corresponding host.

\begin{figure}
	\includegraphics[width=\columnwidth]{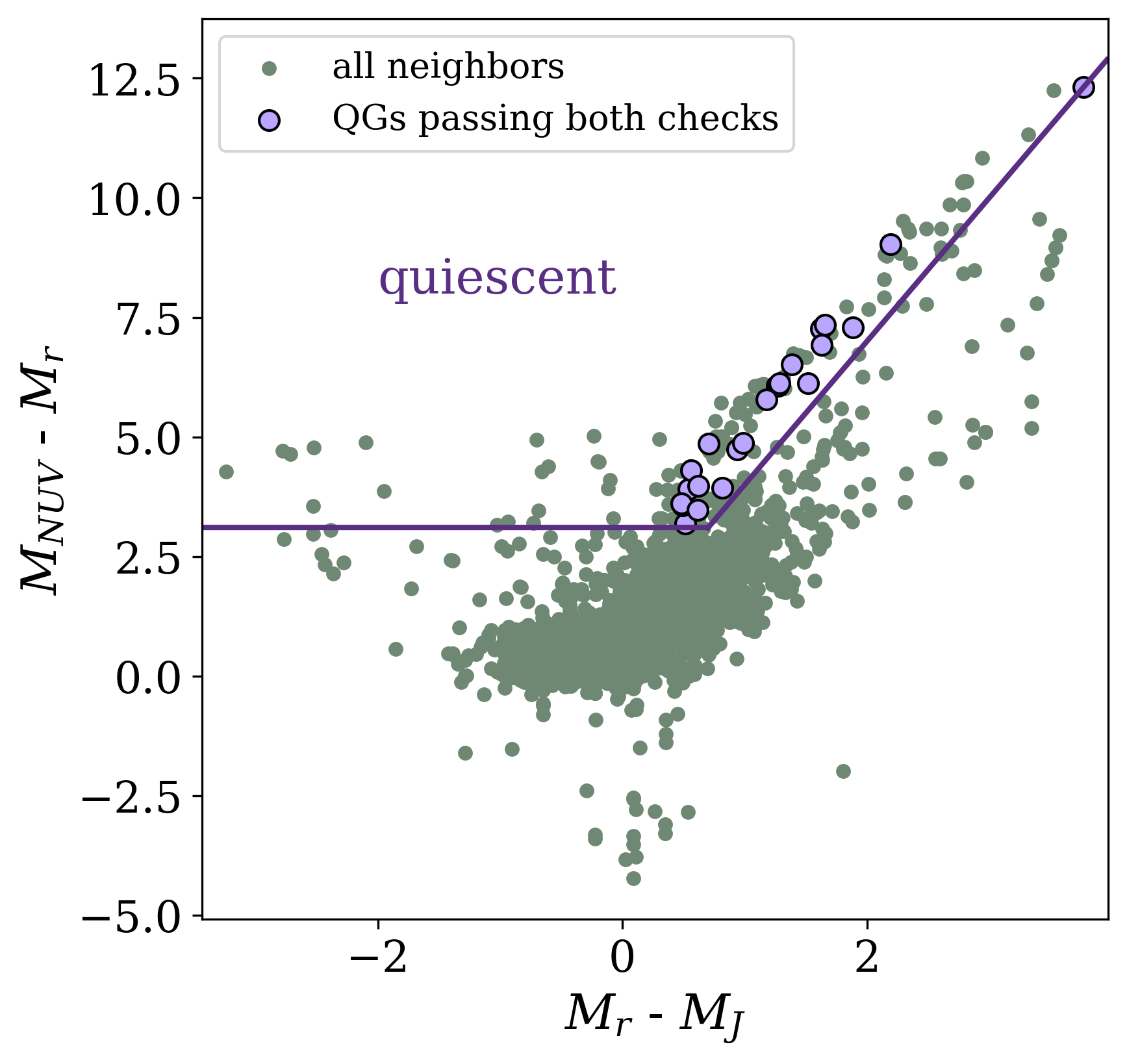}
    \caption{NUV-r-J diagram for neighbors of the S$\mathrm{\hat{O}}$NG sample. The purple line delineates the QG region from \cite{Ilbert2013}. The lavender dots denote the neighbors that pass both this threshold and our sSFR criterion.}
    \label{fig:NUV-r-J}
\end{figure}

Adopting $\mathrm{log_{10}(sSFR/yr)}$ $<-9.8$ as the criterion for post-starbursts and quiescents, we find that only 5\% of our neighbors meet this criterion. If anything, their high sSFRs would put them in the actively star-forming category. This is seen in Fig.~\ref{fig:mstar_vs_sfr}, where their $\mathrm{M_*}$ and SFR fall mostly into the main-sequence region parameterized by \cite{Rinaldi2022}. Nearly half of the neighbors (enclosed by the 0.5 contour level in Fig.~\ref{fig:mstar_vs_sfr}) lie between the main sequence and our post-starburst/quiescent threshold, indicating a transitional demographic, comparable to the ``green valley'' at low-$z$ \citep[e.g.,][]{Wyder2007, Belfiore2017}. Regardless, we note that in the absence of spectroscopy or additional medium band photometry, SED fits can produce inaccurate SFR and $\mathrm{M_*}$ results due to, for example, the age-dust-metallicity degeneracy, or assumptions about nebular emission \citep{Choe2025, Jones2025}. As a complementary test, we apply the NUV-r-J rest-frame color-color diagram method from \cite{Ilbert2013}, using NUV, r, and J absolute magnitudes from the COSMOS2025 \texttt{LePHARE} catalog \citep{Shuntov2025}. Only 7.4\% of our neighbors pass this test; Fig.~\ref{fig:NUV-r-J} shows them as green dots above the purple line. They exhibit a 15\% overlap with the neighbors that passed the sSFR check; this overlap is shown as lavender dots in Fig.~\ref{fig:NUV-r-J}.

Our analysis has found low quenched fractions in our sample ($\approx 5-$7\%). Regardless, as a precaution, we cannot rule out that at least \textit{some} of these QGs are dusty SFGs misidentified by the SED-fitting software, a phenomenon already noted by previous works \citep[e.g.,][]{AntwiDanso2023, Forrest2024}. Such errors can impact not only the sSFR check, but also the NUV-r-J inspection, which has historically misclassified dusty SFGs and compact young systems as they occasionally scatter into the NUV-r-J quiescent space at $z$ $>$ 3 \citep[e.g.,][]{Hwang2021}. In tandem, these potential sources of over-counts suggest that our true quenched fractions may be even lower than reported.

From a cross-comparative perspective, our results differ greatly from previous works that have probed up to $z<2$ \citep[e.g.,][]{Hartley2015, Treyer2018} and found that the neighbors of QGs show higher quenched fractions, i.e. galactic conformity. It also differs from a new analysis by \cite{2025ApJ...978...17M} using the MAGAZ3NE survey \citep{Forrest2020b}, which shows elevated quenched fractions in protoclusters around $z$ $\sim$ 3 ultra-massive quiescent centrals (M$_*$ $>$ 10$^{11}$ M$_\odot$). However, these results are not contradictory, but complementary in the grand scheme of mass-dependent environmental quenching. When it comes to this type of quenching, we know that there are many possible pathways. Firstly, when a low-mass galaxy enters a dense environment, the exerted ram pressure can cause gas depletion, either by quickly \textit{stripping} the star-forming gas \citep[e.g.,][]{Abadi1999} or by cutting off its coronal gas supply, thus gently quenching the galaxy in what we will refer to as \textit{desiccation}\footnote{\label{term-change}In the literature, the words `strangulation' and `harassment' are often evoked to describe these processes, respectively. But such terms were borrowed from violent human actions with quick, irreversible effects---not at all analogous or accurate to the galaxy interactions in question. Moreover, they are very recent coinages: `strangulation' does not appear in the papers that first noted the process \citep{Larson1980, Balogh2000}, and neither does `harassment' in the equivalent work by \cite{Farouki1981}. Inspired by an article by \cite{madrid2024} and discussions with our co-authors, we agree that these terms need changing. We instead adopt `desiccation' and `erosion' from geography; `desiccation' refers to the drying-up of a body of water over time, while `erosion' describes how soil is gradually removed via high-speed impacts from adjacent bodies of water. We consider these new terms to be more appropriate representations of the galactic processes, since astronomy has already described gas-rich galaxies as `wet', and the analogy between galaxies and bodies of water has cultural and linguistic precedents worldwide.} \citep[e.g.,][]{Larson1980, Balogh2000}. Alternatively, the in-falling galaxy can have extreme, repeated tidal interactions with galaxies along its path, resulting in what we call \textit{fly-by erosion}$^{\ref{term-change}}$---whereby the in-falling galaxy loses material and/or changes shape to the point of triggering a starburst that rapidly drains away its gas component \citep{Farouki1981}. Much as these quenching pathways differ, they are unified by the crucial role of mass \citep[e.g.,][]{Hirschmann2014, Mao2022}; a galaxy in a less massive cluster will encounter less mass on the way and quench more slowly, which corresponds to a longer conformity timescale---so long that not enough time may have passed in the early Universe \citep[$\sim1$\,Gyr at $z = 3-6$;][]{Chamberlain2024}. And within a cluster of low-mass galaxies around a central, dominant QG, the QG's mass will govern the timescale required to environmentally quench its neighbors. In the case of \cite{2025ApJ...978...17M}, all the QGs are ultra-massive (M$_*$ $>$ 10$^{11}$ M$_\odot$); thus, the drastic gravitational potential gradient can efficiently trigger stripping, desiccation, or fly-by erosion, leading to more elevated quenched fractions in their neighborhood. But in contrast, most of our S$\mathrm{\hat{O}}$NG QGs have M$_*$ 1 dex lower (see the red dots in Fig.~\ref{fig:mstar_vs_sfr}); this makes the gravitational potential gradient less pronounced and less conducive to environmental quenching. In other words, examining our results in relation to other z $\gtrsim$ 3 studies suggests that in the early Universe, environmental quenching efficiencies, and thus conformity timescales, correlate strongly with the mass of the protocluster central.

\begin{figure*}
	\includegraphics[width=\textwidth]{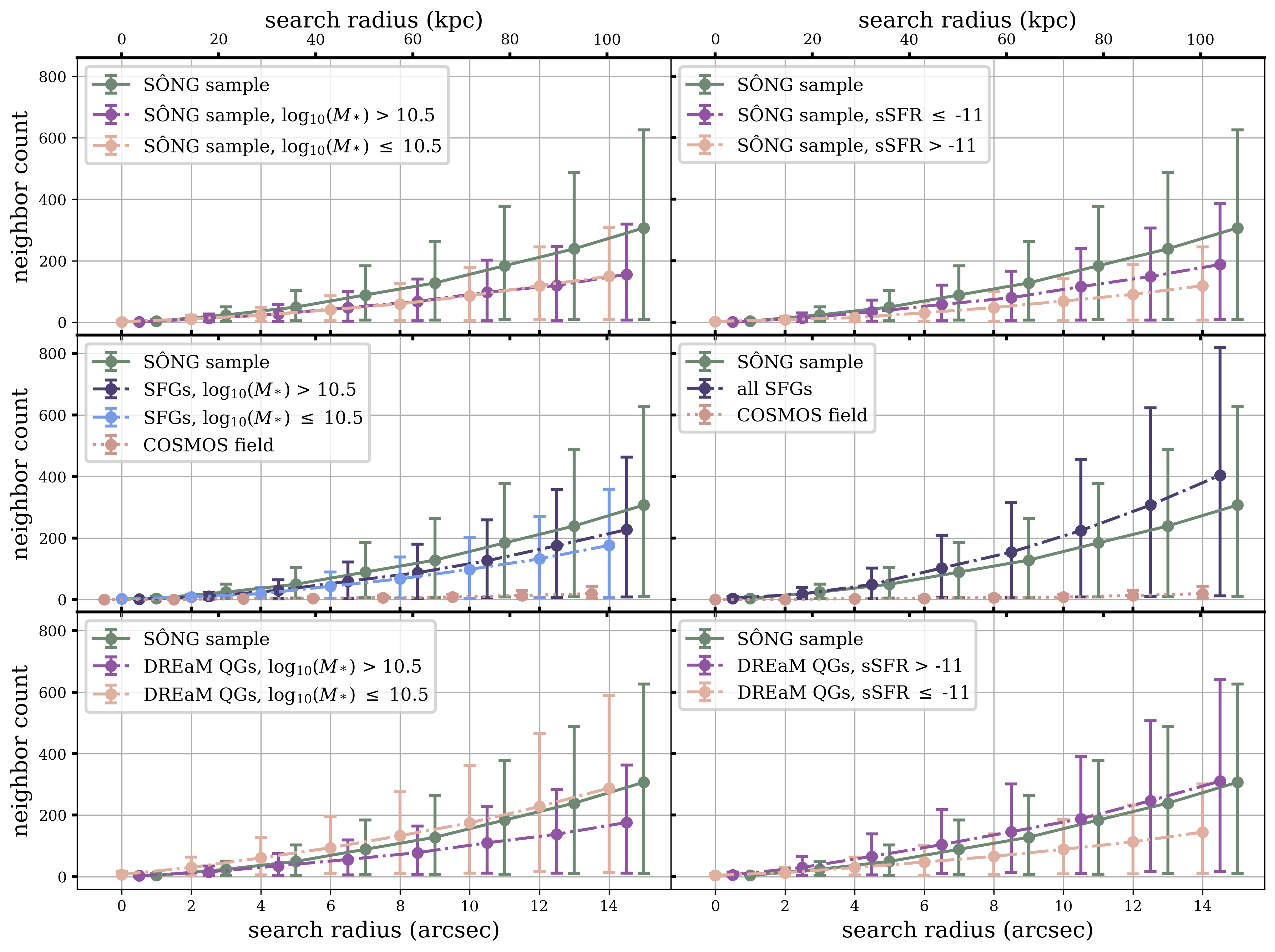}
    \caption{Neighbor count (also called clustering signal or strength) versus annulus for the S$\mathrm{\hat{O}}$NG sample (green), compared to other samples. From left to right, top to bottom: S$\mathrm{\hat{O}}$NG differentiated by $\mathrm{M_*}$ (left plot, top row) and sSFR (right plot, top row); 100 randomly-selected SFGs differentiated by $\mathrm{M_*}$ and 100 random field pointings in COSMOS-Web (left plot, middle row); 100 randomly-selected SFGs and  field pointings from COSMOS-Web (right plot, middle row); 100 random QGs from DREaM differentiated by $\mathrm{M_*}$ (left plot, bottom row) and sSFR (right plot, bottom row). For all the COSMOS data (QGs, mass-matched SFGs, and field pointings), the error bars are the 16$^\mathrm{th}$ and 84$^\mathrm{th}$ percentiles of the counts. The top x-axis represents the annuli converted into kpc at $z\sim3.5$, based on concordant $\Lambda$CDM cosmology assumptions from $\S$\ref{sec:intro}. All samples are analyzed with the same annuli, but their graphs are shifted left by steps of 0.5$''$ for visual clarity, except S$\mathrm{\hat{O}}$NG which is unshifted.}
    \label{fig:centipede}
\end{figure*}

Still, given the uncertainties of quantifying high-$z$ quiescence via SED fitting \citep[e.g.,][]{AntwiDanso2023, Forrest2024}, we must note that our results may change with spectroscopic follow-ups on the same sample. Such efforts will require high-resolution, or at least extensive spectroscopic programs aimed to measure high redshifts like the recent CAPERS \citep[PID 6368;][]{CAPERS}. Other pathways include additional deep mid-IR coverage from surveys with configurations like SMILES \citep{Alberts2024} or medium-band programs like MINERVA \citep{Muzzin2025}.

\subsection{Clustering signals} \label{subsec:clustering}

To quantify clustering while accounting for $z\mathrm{_{phot}}$ uncertainties, we implement a Monte Carlo (MC) approach which samples each neighbor's $\texttt{LePHARE}$ $z\mathrm{_{phot}}$, using its asymmetric $\mathrm{1\sigma}$ confidence intervals to approximate the underlying redshift probability distribution. For each neighbor, we sample 1,000 random $z\mathrm{_{phot}}$ values from its $z\mathrm{_{phot}}$ distribution and count all the times that a sampled $z\mathrm{_{phot}}$ value is within the host's $\mathrm{1\sigma}$ redshift uncertainty. With the satisfactory $z\mathrm{_{phot}}$ values, we then step through concentric annuli of width 1$''$ out to 15$''$ (our original search radius), adding up the number of neighbors that pass per each of the 1,000 iterations. For each annulus, we report the 16$^\mathrm{th}$, 50$^\mathrm{th}$, and 84$^\mathrm{th}$ percentiles of the neighbor counts across the 1,000 MC iterations.

This MC method is performed on the following samples: (1) the 171 S$\mathrm{\hat{O}}$NG QGs; (2) 100 randomly selected, $\mathrm{M_*}$-matched, coeval SFGs with $3 \leq z_\mathrm{phot}<5$; and (3) 100 random pointings in the background COSMOS-Web field. In the third case, as we are quantifying clustering signals around random sets of coordinates that may not always overlap with a real galaxy, there is no host $z\mathrm{_{phot}}$. To adapt our method, we treat the median $z\mathrm{_{phot}}$ of the S$\mathrm{\hat{O}}$NG sample as the $z\mathrm{_{phot}}$ of the centers of all the random pointings. We show the results in Fig.~\ref{fig:centipede}.

Finally, we introduce simulated data for comparison. We pick the semi-empirical Deep Realistic Extragalactic Model \citep[DREaM,][]{DREaM}, which uses subhalo abundance matching to build synthetic galaxy catalogs, and whose results have successfully reproduced galaxy bimodality beyond $z \sim 4$. With DREaM's catalog, we first apply a mass cut, taking only galaxies with $\mathrm{M_*}$ $\geq 10^{8}$ $\mathrm{M_\odot}$ to mimic COSMOS2025 mass completeness conditions \citep{Shuntov2025}. Then, we randomly select 100 QGs at 3 $\leq$ $z$ $<$ 5 with $\mathrm{M_*}$ $> 10^{10.5}$ $\mathrm{M_\odot}$; our quenching criterion is $\mathrm{log_{10}(sSFR)}$ $\leq -9.8$ yr$^{-1}$, mirroring $\S$\ref{subsec:sample-construction}. We then use the same annuli from earlier (1$''$ to 15$''$) to find neighbors for each QG. Within each annulus, we also compute the $z\mathrm{_{phot}}$ median, 16$^\mathrm{th}$, and 84$^\mathrm{th}$ percentiles of the neighbors from $\S$\ref{sec:search}. We use these values to construct a redshift uncertainty distribution for each DREaM neighbor. By sampling that distribution 100 times, we perturb the neighbor's redshift to mimic observational $z\mathrm{_{phot}}$ uncertainties. Lastly, we count all the times that the host redshift is within range of the uncertainties of the perturbed neighbor redshift. The median, 16$^\mathrm{th}$, and 84$^\mathrm{th}$ percentiles of DREaM's neighbor counts are plotted in Fig.~\ref{fig:centipede}.

From Fig.~\ref{fig:centipede}, it is clear that our S$\mathrm{\hat{O}}$NG sample exhibits diverse but visible clustering signals, and there is little difference between subpopulations as divided by $\mathrm{M_*}$ or sSFR. In other words, our QGs are strongly clustered on average, and the clustering does not distinctly favor those with less or more stellar mass, or those that are ``more'' or ``less'' quenched. This is corroborated by the DREaM QGs (Fig.~\ref{fig:centipede}, bottom row), which also exhibit fairly unbiased clustering signals for different mass subsets. The stronger clustering around the subset of DREaM QGs with sSFR $>$ $-$11 is likely due to selection bias, as this population constitutes 75$-$80\% of our randomized DREaM sample. Across all our observed and simulated samples, the wide variance, whose lower limit nears zero, is attributable to less common cases where there are no robust neighbors found around a source.

\begin{figure*}[th!]
  \includegraphics[width=\textwidth]{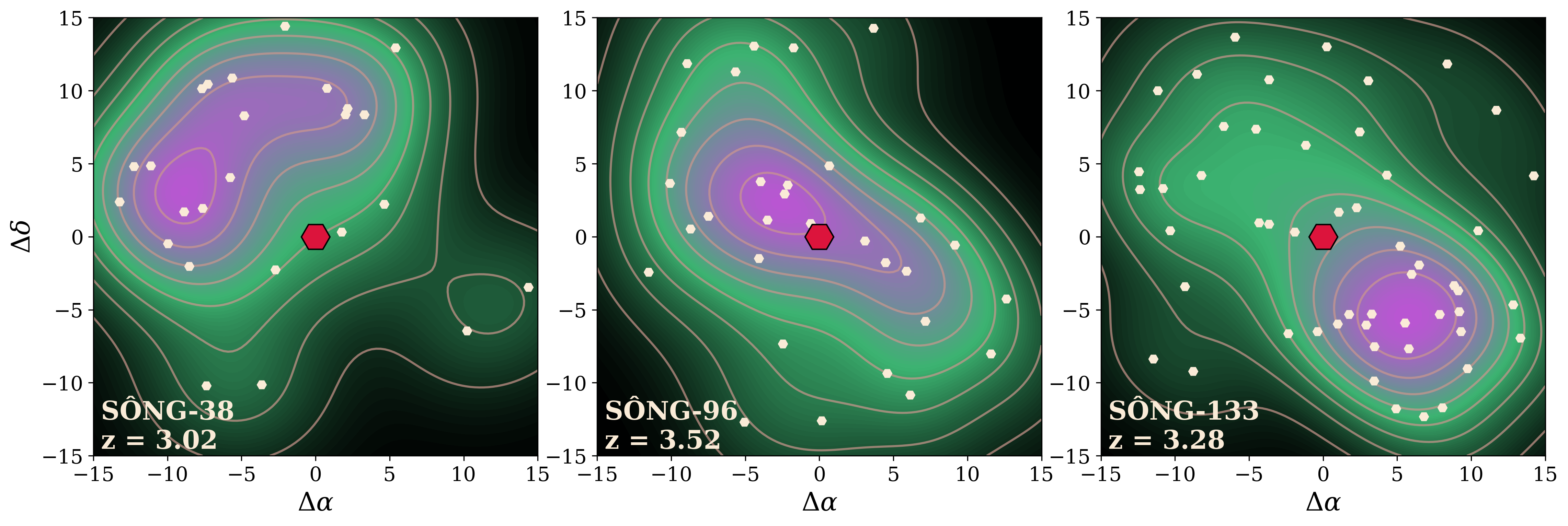}
  \caption{Temperature maps of the three exceptional cases ($q_{P95}$ $\geq$ 2) in \ref{subsec:filament}, showing potential filamentary structures. The large, central crimson hexagon represents the QG, and the smaller ivory hexagons represent its neighbors. Purple represents density regions. Each QG's $z\mathrm{_{phot}}$ is indicated at the bottom left corner of its respective panel.}
  \label{fig:heat_maps}
\end{figure*}

However, the picture gets more complicated if we compare our sample to its original observational catalog. Even though the S$\mathrm{\hat{O}}$NG galaxies do exhibit higher clustering signals than random field pointings (as seen across the middle row of Fig.~\ref{fig:centipede}), such signals are not higher than coeval, stellar-mass-matched SFGs. Conversely, as shown in the middle right panel of Fig.~\ref{fig:centipede}, our sample and the SFGs are almost equal in terms of neighbor count up to 9$''$, then the clustering around SFGs begins to be higher than that of the QGs, yet remains compatible within error bars. The marginal difference continues out to 15$''$, which is 25\% of the r$_\mathrm{vir}$ for a galaxy in this mass range at $z\sim3.5$. If robustly confirmed, these results may suggest a picture of the early Universe where proto-groups and protoclusters with massive quiescent and star-forming centrals exhibit more or less the same neighbor densities---i.e. high-$z$ QGs reside within locally underdense or `equi-dense' environments with respect to coeval, mass-matched SFGs. 

Our results are a strange anomaly in the study of quiescent galaxies; they differ from well-known low-$z$ trends where QGs exhibit higher clustering signals than SFGs \citep[e.g.,][]{Berti2019}. While there are recent JWST studies of high-$z$ QGs that report `overdensities' \citep[e.g.,][]{Helton2024, deGraaff2025}, we must note that they are describing overdensities with respect to the average density of the galaxy population at their epochs. Our results do not challenge these works, but complement the picture. Indeed, high-$z$ QGs do show more clustering than average, and this is clearly corroborated by our field comparison; still, such clustering is not higher than coeval SFGs of similar $\mathrm{M_*}$. That second, complementary comparison of QGs and SFGs is not present in the aforementioned recent studies.

But what do these results mean for the study of QGs? As noted in $\S$\ref{sec:intro}, low-$z$ QGs are most often found in local overdensities, which is attributed to them being embedded in more massive DM halos, thus gravitationally attracting more neighbors. If the lack of higher clustering signals in our high-$z$ sample persists even when the clustering amplitudes and differences between our subsamples are more strongly constrained, this could indicate that the aforementioned low-$z$ clustering trends are not always reproducible at high-$z$, and instead may take several Gyr to develop. Indeed, recent semi-analytical models and cosmological simulations are finding that high-$z$ QGs do not strictly reside in local overdensities, but a diversity of environments, including filament knots and void-like underdensities \citep[e.g.,][]{DeLucia2025}. A new study by \cite{Kimmig2025} using the Magneticum Pathfinder simulations also outlines a quenching pathway whereby a massive galaxy can only quench at $z$ $>$ 3 if its environment becomes a local underdensity, which then weakens the feeding filaments and makes the gas component more prone to disruption and ejection by AGN feedback. Collectively, these models are unveiling a convoluted early Universe where massive QGs do not always live in overdensities, suggesting that the massive DM halos associated with their low-$z$ counterparts may still be fledgling and have yet to stabilize. If confirmed, our results may be the first-ever observational evidence of such theoretical environmental diversity. More robust methods are needed to definitively assess these high-$z$ clustering trends.

\subsection{Filament identification} \label{subsec:filament}

Many recent spectrophotometric studies have detected filamentary structures in protoclusters at $z\geq 2$ \citep[e.g.,][]{Umehata2019, Shi2021, Tanaka2024}. Lacking spectroscopic data for our sample, we suggest an alternative geometric method to identify possible filamentary structures and quantify their directionality. A brief overview is provided below.

Our method relies on the on-sky 2D circle around the central QG produced by the search radius in $\S$\ref{angular}. We break this 2D circle into equal sectors and introduce the sectoral number density $n_s$, which is expressed as:

\begin{equation}
\label{eq:sectoral}
    n_s = \frac{N_{HX} q}{\pi r^2}
\end{equation}

\noindent where $N_{HX}$ is the total neighbor count in each sector, $q$ is the number of sectors, and $r$ is the search radius. $n_s$ is compared against $n_f$, the number density of neighbors in the general coeval field. To ensure statistical robustness, we will consider the 50$^\mathrm{th}$ (median), 84$^\mathrm{th}$, and 95$^\mathrm{th}$ percentiles of the coeval field's number of neighbors per arcsec$^2$ as the threshold for comparison. If a protocluster has at least one sector with $n_s$ $>$ $n_f$ across all three corresponding values of $n_f$, that protocluster hosts candidate overdensities consistent with potential filaments, i.e. filamentary signals.

For our analysis, we apply equal radial sextants ($q$ $=$ 6) centered on the host QG, with $r$ $=$ 15$''$. We further define a rotation angle, $\alpha$, that is 12.5\% of the angle subtended by each sector, and rotate the sextants together by within $[-2\alpha, 2\alpha]$ to fully encapsulate potential cross-sector structures, while constraining the overlap of $N_{HX}$ between sectors that could contaminate directionality. After deriving $n_s$ for all the sextants across five rotations, we compute the average $n_s$ of each sextant and compare this to $n_f$. In our case, $n_f$ corresponds to the neighbor count within 15$''$ of the random pointings from $\S$\ref{subsec:clustering}, divided by the area of the search circle, which is $\pi$ $\times$ (15 arcsec)$^2$ $=$ 707 arcsec$^2$ ($\sim$ 0.2 arcmin$^2$). We report the $n_f$ 50$^\mathrm{th}$, 84$^\mathrm{th}$, and 95$^\mathrm{th}$ percentiles to be 0.0438, 0.0509, and 0.0552 neighbors/arcsec$^2$, respectively.

\begin{table}[b]
    \hspace*{-0.5cm} 
    \begin{tabular}{@{}lcccr@{}}
        \toprule
        & $q \geq 1$ & $q \geq 2$ & $q \geq 3$ & in AMICO \\
        \midrule
        P50 & 27 & 8 & 3 & 24 \\
        P84 & 12 & 4 & 1 & 12 \\
        P95 & 7 & 3 & 1 & 7 \\
        \bottomrule
    \end{tabular}
    \caption{\label{tab:filaments}Number of QGs from the S$\mathrm{\hat{O}}$NG sample that, from left to right, have $\geq$ 1, 2, and 3 sextants showing filamentary signals; and are identified as potential group members in the AMICO catalog \citep{Toni2025}. The rows are the percentiles used to derive $n_f$.}
\end{table}

The results are summarized in Table \ref{tab:filaments}. Overall, only seven QGs, or 4\% of the S$\mathrm{\hat{O}}$NG sample, make the most robust cut for $\ge 1$ filamentary signal ($q_{P95}$ $\geq$ 1). Three QGs stand out for at least two sextants with $n_s$ $>$ $n_f$, and, in the case of S$\mathrm{\hat{O}}$NG-133, five sextants! Fig.~\ref{fig:heat_maps} demonstrates the possible structures around these three QGs with a visual heatmap. S$\mathrm{\hat{O}}$NG-133 appears to reside between two distinct northwestern and southeastern filament-like overdensities, with the latter comprising more concentrated neighbors. \texttt{CIGALE} estimates reveal that all of this QG's neighbors are inferior in $\mathrm{M_*}$, including $\sim 90\%$ with mass ratios $<$ 1:10; thus, its status as a protocluster central is further supported. Lastly, we compare our most robust cut ($q_{P95}$ $\geq$ 1) to potential galaxy group/protocluster core members in the AMICO catalog \citep{Toni2025} and recover a 100\% match (see Table \ref{tab:filaments}); this includes the three exceptional QGs. We also use the $p$ $\geq$ 0.5 cut for AMICO robustness, and only two QGs from our most robust cut ($q_{P95}$ $\geq$ 1) pass this second cut. Interestingly, these do not overlap with the three exceptional QGs in Fig.~\ref{fig:heat_maps}, but are two distinct QGs: S$\mathrm{\hat{O}}$NG-62 ($z$ $=$ 3.46) and S$\mathrm{\hat{O}}$NG-135 ($z$ $=$ 3.38). In other words, of the QGs which pass our most robust cut ($q_{P95}$ $\geq$ 1), all seven are potential group members in AMICO; three of them exhibit filamentary signals in at least two sectors ($q_{P95}$ $\geq$ 2), while two others pass a second robust cut ($p$ $\geq$ 0.5) in AMICO. Such agreements with AMICO further confirm that these seven QGs reside within overdense environments relative to the field.

Our low fraction of protoclusters exhibiting filamentary signals (4\%) is likely a conservative result, since our method only identifies such signals if at least one sector has $n_s$ $>$ $n_f$. However, it does offer interesting implications for these systems. A mixture of both observations and cosmological simulations have shown that at high-$z$, galaxies in filaments can have high SFRs due to continual cold gas accretion as they travel along the filaments \citep{Woods2014, Umehata2019, Waterval2025}. In this regard, it is plausible that QGs like S$\mathrm{\hat{O}}$NG-133, which show filamentary signals and are neighbored by mostly low-mass SFGs, will receive enough cold gas over time to reignite star formation, i.e. `rejuvenate'. Still, we note that such future predictions are only conjectural. As it stands, our method's 2D nature prevents distinguishing real filamentary structures from chance projections or group asymmetries. In addition, it is only sensitive to transverse filamentary structures, while line-of-sight filaments remain unquantifiable given $z\mathrm{_{phot}}$ uncertainties. A precise, extensive study of filaments will require both refined spectroscopic data---the kind only obtainable with instruments like NIRSpec PRISM---and algorithms that can diagnose/reconstruct large-scale structures. Our sector-based method is an effective first step to identify extraordinary galaxies whose anisotropic neighbor distributions warrant such observational and computational follow-ups.

\section{SPECTROSCOPIC SAMPLE} \label{sec:specz}

\begin{table*}[t]
\centering 
\hspace*{-1.8cm} 
\begin{tabular}{llllll}
\hline
Galaxy & RA & Dec & $z\mathrm{_{spec}}$ & M$_*$ & References \\
& [deg.] & [deg.] & & [$\times$ 10$^{11}$ M$_\odot$] & \\
\hline
\object{ZF-COS-20115}         & 150.06149 & +2.37868 & 3.715 & 1.24 & \cite{Glazebrook2017}\\
\object{ZF-COS-18842}         &   150.08728 & +2.3960431 &  3.782 & 0.40 & \cite{Schreiber}\\
\object{ZF-COS-19589}         &   150.06671 & +2.3823645  &  3.715 & 0.64 & \cite{Schreiber}\\
\object{Huong-Giang (COS55-129098)} & 150.43732 & +2.463920 & 3.336 & 0.19 & \cite{AntwiDanso2025}\\
\object{Thu-Bon (COS55-128636)} & 150.45459 &  +2.455994 & 3.757 & 0.72 & \cite{AntwiDanso2025}\\
\object{Hong} & 150.108873 & +2.330539 &   3.238 & 0.07 & This work \\
\object{Saigon} & 150.106932 & +2.377961 & 4.550 & 0.01 & This work \\
\hline
\end{tabular}
\caption{\label{Table:speczgalaxysample} The sample of seven spectroscopically confirmed massive QGs. In the Galaxy column are the names of these galaxies and/or their catalog IDs in previous works if applicable; the naming is explained in detail in $\S$\ref{subsec:specz_nbr_match}. The $\mathrm{M_*}$ for Hong and Saigon are independently measured in this work; for the remaining galaxies, their RA, Dec, z$_{spec}$ and M$_*$ come from \cite{Suzuki2022} and \cite{AntwiDanso2025}.}
\end{table*}

Motivated by the results of our photometric study, we also apply our methodology to a number of spectroscopically confirmed QGs in the COSMOS field. This sample consists of seven QGs, with $3<\mathrm{z_{spec}}\leq4.5$, and $\mathrm{M_*}$ $\sim$ $10^9-10^{11}$ $\mathrm{M_\odot}$. More details are provided in Table \ref{Table:speczgalaxysample}.

\begin{figure}[t]
	\includegraphics[width=\columnwidth]{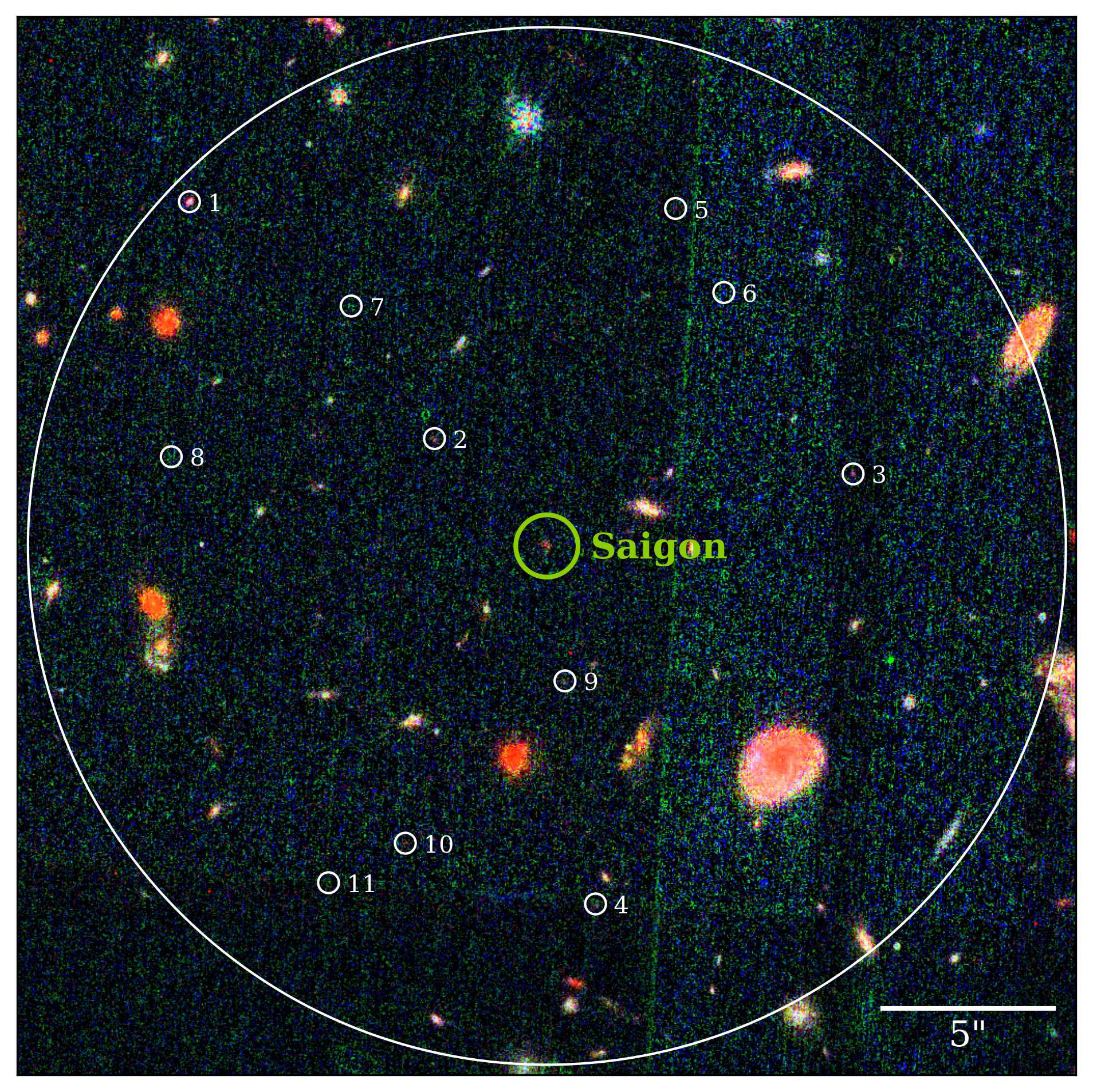}
    \caption{RGB image of Saigon (circled green), a spectroscopically confirmed QG at $z$ = 4.55. All 11 neighbors, which constitute the larger Saigon protocluster, are circled white, their names abbreviated as numbers on their right. The filters used for R, G, and B are the same as Fig.~\ref{fig:SONG-154_2figs}.}
    \label{fig:Saigon}
\end{figure}

Five of our spectroscopically-confirmed QGs originate from previous works. Two of these were identified in the FENIKS survey and studied with Keck-MOSFIRE by \cite{AntwiDanso2025}. The remaining three were selected from the ZFOURGE survey and confirmed via Keck-MOSFIRE K-band spectra \citep{Glazebrook2017, Schreiber}, then studied with ALMA by \cite{Suzuki2022}. 

Lastly, our sample also introduces two new QGs, identified in our work using the DAWN JWST Archive \citep{Valentino2023} and measured spectroscopically in a JWST Director's Discretionary Time (DDT) program \citep{CoulterDDT}. These are named in Table \ref{Table:speczgalaxysample} as Hong and Saigon. Hong is classified as a QG based on its PRISM spectrum (see Appendix \ref{spectra_appendix}), which shows a 4000\r{A} break at D$_{n}$4000 $=$ 1.34, and UVJ colors that fall into the \cite{Belli2019} extended post-starburst region for high-$z$. Saigon's classification is detailed in $\S$\ref{subsec:Saigon}.

\subsection{Catalog matching and neighbor search} \label{subsec:specz_nbr_match}

With the coordinates of our seven QGs, we begin by matching them in COSMOS2025. Even though we successfully retrieve all seven from the catalog, only three (Huong-Giang, Thu-Bon, and ZF-COS-18842) have the sufficient $z\mathrm{_{phot}}$, $\mathrm{M_*}$, and sSFR that qualify them as high-$z$ massive QGs based on the criteria in $\S$\ref{subsec:sample-construction}. For the four disqualified galaxies, the reasons are multi-layered. ZF-COS-20115 has $z\mathrm{_{phot}=0.9761}$ in the catalog; however, its spectroscopic redshift has long been attested in the literature as $z\mathrm{_{spec}}=3.715$ \citep{Schreiber, Suzuki2022}. Additionally, two of the disqualified QGs (ZF-COS-20115 and ZF-COS-19589) lack both SFR and $\mathrm{M_*}$ estimates in the \texttt{CIGALE} catalog \citep{ArangoToro2025}, while the remaining two (Hong and Saigon) have \texttt{CIGALE} sSFR $=$ $-$9.46 and $-$9.04 yr$^{-1}$, respectively, suggesting that the SED fitting software preferred a dusty starburst SED when only using photometric data. These inconsistencies signal a potential risk in our previous reliance on SED fits. We move forward relying on the spectroscopic data, treating the seven galaxies as QGs even if their properties in COSMOS2025 fail our criteria in $\S$\ref{subsec:sample-construction}.

Our neighbor search around the spectroscopically confirmed sample is conducted similar to $\S$\ref{sec:search}, with the only difference being the redshift step, which now compares a single $z\mathrm{_{spec}}$ value of the host to the 1$\sigma$ range of each neighboring source's $z\mathrm{_{phot}}$. In the end, we cannot identify any sources around ZF-COS-18842, ZF-COS-19589, and ZF-COS-20115 that meet all our neighbor criteria. While indeed, all seven QGs are found with neighboring sources which pass our first three checks (angular separation, significant detections, photometric redshift), no sources remain around the three aforementioned QGs after being compared to brown dwarf SED fits described in $\S$\ref{subsec:zSED}. This differs from findings from previous studies that ZF-COS-20115 has a dusty neighbor at around the same $z\mathrm{_{spec}}$, making up a so-called `Jekyll and Hyde' pair \citep{Schreiber, Suzuki2022, Liu2025}. A perusal of COSMOS2025 reveals that Hyde, the missing dusty neighbor, does not pass any of our criteria only because the COSMOS-Web segmentation does not classify it as a distinct galaxy from Jekyll (ZF-COS-20115), which indicates that our original photometric catalog is missing neighbors that are blended with their hosts. Lastly, we crosscheck with the AMICO catalog \citep{Toni2025}, which is based on the same SED-fitting outputs as ours. Among our seven-galaxy sample, AMICO does not report ZF-COS-20115 nor ZF-COS-19589 as potential group members either, but does include ZF-COS-18842.

The four other QGs all have neighbors; this includes Huong-Giang and Thu-Bon, whose COSMOS2025 physical parameters meet our definition of massive QGs (see $\S$\ref{subsec:sample-construction}). We crosscheck with AMICO and also find three of these four QGs as potential group members; the only exception, Saigon, may be due to AMICO's redshift cut at $z$ $=$ 3.7. To ensure the robustness of the final sample, we also visually inspect the sources that pass our checks and eliminate those resembling stellar spokes ($n=2$ removed). In the end, the four QGs are found with a sum total of 36 neighbors. For ease of discussion, we name only these QGs, and thus their associated protoclusters, after major rivers of Vietnam, as seen in Table \ref{Table:speczgalaxysample}.\footnote{Our nomenclature is both a thematic continuation from $\S$\ref{subsec:sample-construction}, and an homage to the roughly 2,300 rivers of Vietnam, most of which are red and ``heavy'' with silt, analogous to our high-$z$ QGs.}

\subsection{Protocluster properties}

We identify 36 neighbors around our four QGs, with median mass $\mathrm{M_*}$ $=$ 1.67 $\times$ $10^8$ $\mathrm{M_\odot}$. None of these neighbors have corresponding $z\mathrm{_{spec}}$ in a recently released compilation of spectroscopic catalogs for the COSMOS field by \cite{Khostovan2025}. With regard to host-neighbor relations, Huong-Giang, Thu-Bon, and Hong dominate in $\mathrm{M_*}$ compared to their neighbors; this, based on the QGs' $z\mathrm{_{spec}}$ between 3.238 and 3.757, makes them some of the earliest known protocluster cores with quiescent centrals in COSMOS. Saigon has 11 neighbors, which makes its protocluster the earliest and most abundant of our sample; still, Saigon's $\mathrm{M_*}$ is only $\sim$ 31\% that of the neighboring Saigon HX8 (labeled in Fig.~\ref{fig:Saigon}), suggesting it may not be the protocluster central. A more complete account of this peculiar system is given in $\S$\ref{subsec:Saigon}.

\begin{figure*}[t]
  \includegraphics[width=\textwidth]{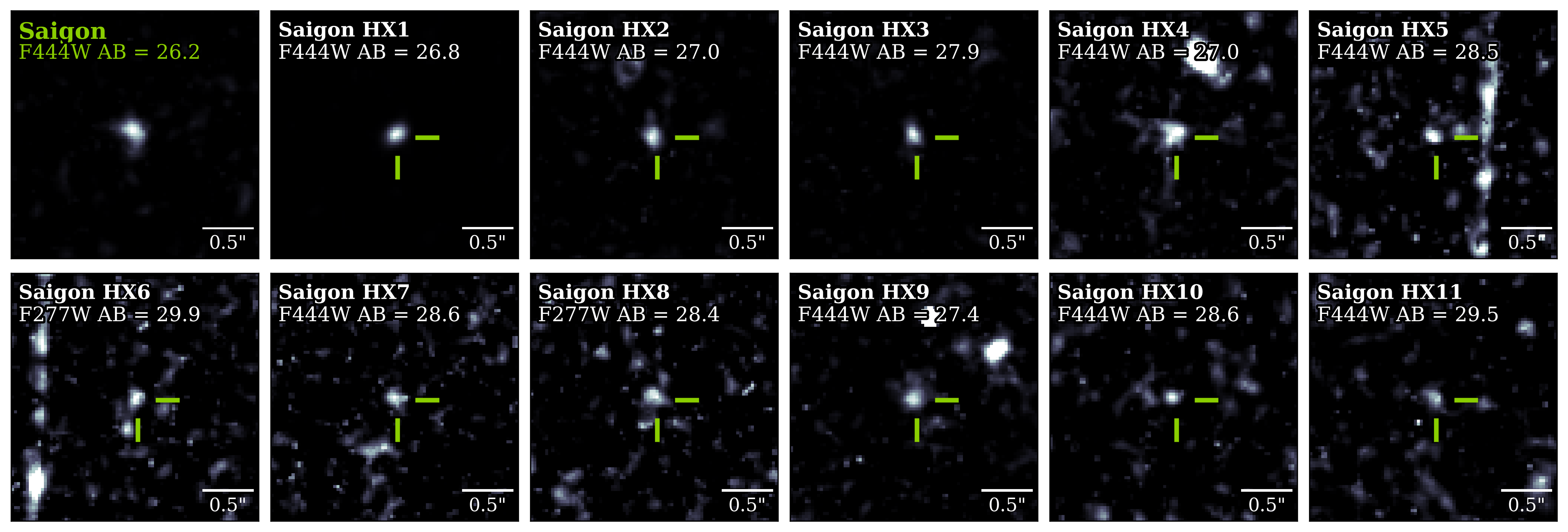}
  \caption{$\chi^2$ detection images of Saigon ($z$ $=$ 4.55) and its neighbors, constructed and reduced from all four JWST NIRCam bands (F115W, F150W, F277W, and F444W). The image contrast has been manually adjusted for visibility. Each galaxy's AB magnitude in the reddest band available is annotated on the top left corner of its corresponding cutout.}
  \label{fig:Saigon_nbrs}
\end{figure*}

Similar to $\S$\ref{subsec:prop}, we find low quenched fractions among these neighbors, with only 6\% passing the $\mathrm{log_{10}(sSFR)}$ $<$ $-9.8$ yr$^{-1}$ criterion. Consistent with the S$\mathrm{\hat{O}}$NG sample, conformity does not seem to have taken hold at this high-$z$ regime. Replicating the sectoral method in $\S$\ref{subsec:filament}, we cannot find any filamentary signals; thus, it is impossible to speculate whether these QGs may reignite star formation later on. Further high-resolution spectroscopic surveys and deep-field observations will be necessary to examine all four protoclusters in more detail. 

\subsection{Saigon, a noisy protocluster at z = 4.55} \label{subsec:Saigon}

Saigon, named after a river in today's southern Vietnam, is one of the two high-$z$ QGs that are first spectroscopically confirmed in this work. Its spectrum is shown in Appendix \ref{spectra_appendix}, while an RGB image of its protocluster can be found in Fig.~\ref{fig:Saigon}. Saigon is remarkable for being at $z\mathrm{_{spec}}$ $=$ 4.55, the highest among our sample. This high redshift is well beyond the $z$ $=$ 3.7 cutoff of the AMICO catalog \citep{Toni2025}, which explains why it is not included there as a potential group member.

Saigon was first identified serendipitously in a JWST DDT program \citep[PID 6585;][]{CoulterDDT} to look for high-$z$ supernovae. Its resolved PRISM spectrum is made publicly available via the DAWN JWST Archive \citep{Valentino2023}. Given Saigon's high redshift and faint signals (SNR $=$ 4.85), it is difficult to pinpoint its evolutionary stage by color, whether through the traditional UVJ analysis or innovative methods like \textit{$(ugi)_s$} \citep{AntwiDanso2023}. Nevertheless, we are still able to categorize it as a QG. This is thanks to its strong Balmer break typical of $z$ $>$ 3 post-starburst/quiescent galaxies \citep{DEugenio2020}, and an overall lack of strong emission lines except H$\alpha$ and [O \textsc{iii}] $\lambda \lambda$ 4959, 5007\r{A} (see Fig.~\ref{fig:two_spectra}), which in tandem suggests a high-$z$ post-starburst galaxy rather than a dust-obscured SFG. Particularly, [O \textsc{iii}] $\lambda\lambda$ 4959, 5007\r{A} has a moderate luminosity of 8.51 $\times$ 10$^{40}$ erg s$^{-1}$, indicating faint AGN activity. In general, Saigon's spectroscopic profile reveals a fairly young galaxy that has recently undergone a starburst, possibly induced by AGN. Using $\texttt{CIGALE}$ outputs from \cite{ArangoToro2025}, we estimate $\mathrm{M_*}$ $= 1.33 \times10^9$ $\mathrm{M_\odot}$ for Saigon, making it the least massive QG of our sample. QGs of this size ($\sim10^9-10^{10}$ $\mathrm{M_\odot}$) are categorized as intermediate-mass in the recent literature, and have been uncovered in abundance at high-$z$ since JWST's first light \citep[e.g.,][]{Marchesini2023, Sato2024}, but Saigon is the most distant low-mass quiescent galaxy discovered to date.

Fig.~\ref{fig:Saigon_nbrs} is a zoom-in visualization of Saigon's 11 neighbors. None of these neighbors satisfy our criterion for quiescence; their sSFR range is $-$8.14 $\geq$ $\mathrm{log_{10}(sSFR)}$ $\geq$ $-$9.21, falling squarely within the star-forming (and, in some cases, starburst) regime. Five neighbors have mass ratios $\leq$ 1:10 compared to Saigon; five others have mass ratios between 1:10 and 2:5. The last neighbor, HX8, stands out with $\mathrm{M_*}=$ 9.3 $\times$ $10^9$ $\mathrm{M_\odot}$, three times higher than Saigon, and sSFR $=-$8.73 yr$^{-1}$, which indicates a starburst galaxy. We recall that 16\% of our photometric QGs have one more massive neighbor (see $\S$\ref{subsec:prop}); perusing that subset, we recover three cases where a QG has a neighbor with mass ratio $\sim$ 2:1 and sSFR $\geq$ $-$8.7 yr$^{-1}$, suggesting a possible quiescent-starburst pair analogous to Saigon and its HX8. But the closest analog is S$\mathrm{\hat{O}}$NG-72 ($z\mathrm{_{phot}}=$ 4.46; $\mathrm{M_*}=$ 3.36 $\times$ $10^{10}$ $\mathrm{M_\odot}$); its 18 neighbors include HX7 with twice its $\mathrm{M_*}$ and sSFR $=-$8.52 yr$^{-1}$. If confirmed, both Saigon and this analog would exemplify a new `Saigonese' type of high-$z$ protocluster anchored by a near-equal-mass quiescent-starburst pair. For now, we note with caution that the \texttt{LePHARE} $z\mathrm{_{phot}}$ of HX8 is $4.99_{-4.09}^{+5.68}$; the wide variance could mean that it is an unrelated, misidentified source. The same can be said for S$\mathrm{\hat{O}}$NG-72 HX7, which has $z\mathrm{_{phot}}$ $=$ $4.67_{-4.10}^{+0.05}$.

Our preliminary results for Saigon are consistent with spectroscopic studies, which are finding more and more early protoclusters populated by low-mass SFGs \citep[e.g.,][]{Morishita2023, Helton2024}. Nevertheless, Saigon stands out for being a massive, inactive galaxy nestled amid much less massive, yet far more active neighbors. Interestingly, this bimodal contrast has also been corroborated in other high-$z$ protocluster environments, from massive QGs residing amid low-mass SFGs \citep[e.g.,][]{Kakimoto2024} to massive SFGs surrounded by low-mass neighbors with elevated star formation rates \citep[e.g.,][]{Champagne2025}. The lesser activeness of these central galaxies, being `quiet in a world full of noise', suggests an earlier period of rapid mass assembly in the densest, innermost region of a protocluster, which likely overlaps with a filamentary node \citep{Champagne2025}; as a result, these node galaxies became massive and quenched faster than their nodeless neighborhood. Indeed, in the case of Saigon, we cannot definitively say that it is in fact the protocluster central, given the uncertain association of HX8; still, it is undeniable that we are uncovering one of the earliest protocluster nodes known to date that are anchored by a quiescent galaxy. Based on the $\mathrm{M_*}$ of Saigon and all its 11 neighbors, we apply the \cite{ShuntovSHMR} stellar-to-halo mass relations and derive a total DM halo mass $\sim$ $10^{11}$ $\mathrm{M_\odot}$, which sits at the lower end of the halo mass range for protocluster cores to start quenching at $z$ $\sim$ 5 \citep{Chiang2017}. Saigon therefore resembles a recently spectroscopically confirmed massive QG in COSMOS by \cite{Kakimoto2024} at $z=4.53$, which is also surrounded by less massive SFGs and theorized to have become quenched at $z$ $\sim$ 5. Still, it must be noted that Saigon is at a marginally higher redshift ($z\mathrm{_{spec}}=$ 4.55), and its $\mathrm{M_*}$ is 1 dex below the \cite{Kakimoto2024} quiescent central. We propose a hypothetical scenario whereby Saigon quenched rapidly between $z$ $\sim$ 5 and the observed redshift due to high gas consumption, likely induced by a starburst. This would correlate with the low [O \textsc{iii}] luminosity detected.

As it stands, Saigon is the most distant low-mass quiescent galaxy known to date, and is also one of the earliest spectroscopically confirmed quiescent galaxies found in an overdensity. To verify its uniqueness, particularly with complications from HX8, a spectroscopic follow-up is necessary. We note that such programs can be costly, given the faintness of Saigon and its neighbors in the NIRCam F444W and F277W bands (see Fig.~\ref{fig:Saigon_nbrs}). Rough calculations with the JWST Exposure Time Calculator show that even a NIRSpec Multi-Object Spectroscopy (MOS) program reaching the limit of the fastest readout pattern (NRSIRS2RAPID; 1024 integrations per group) will not be able to obtain SNR $\geq$ 1.5 for these sources. Future studies will have to offload the observing cost by both diversifying their instruments and devising more qualitative modeling methods.

\section{SUMMARY \& CONCLUSIONS} \label{sec:sumcon}

In this work, we investigate the local environment of $z >$ 3 massive QGs. This environment is crucial for the profusion of intergalactic phenomena observed for low-$z$ QGs in overdense environments, including: minor mergers that can alter the color gradients and star formation rates of the host; conformity whereby neighbors of QGs show elevated quenched fractions; and higher clustering signals that have implications for galaxy evolution.

With new JWST NIRCam data from the COSMOS-Web program \citep{COSMOSWeb2023, Shuntov2025}, we compile 3 $\leq$ $z_\mathrm{phot}$ $<$ 5 QGs with $\mathrm{M_*}$ $\geq$ $10^{10}$ $\mathrm{M_\odot}$, $\mathrm{log_{10}(sSFR)}$ $\leq -9.8$ yr$^{-1}$, and robust SED-fitting statistics ($\chi^2$ $\leq$ 10). The result is the S$\mathrm{\hat{O}}$NG sample with 171 QGs, around which we look for robust neighbor candidates. We use a mix of observational and numerical methods to eliminate sources: (i) angular distance (15$''\sim$ 100 kpc); (ii) minimum of three significant (SNR $>$ 3) detections; (iii) photometric redshift uncertainty ($\mathrm{|z_{host}-z_{phot}| < 1\sigma}$); and (iv) statistical comparison of SED fits. The main findings are as follows:

\begin{itemize}
    \item We find a total of 2,048 neighbors, $\sim$63\% of which have mass ratio $< 1$:100 compared to their hosts. Our results complement recent JWST searches for low-mass neighbors \citep[e.g.,][]{Suess2023} and reaffirm JWST's auspicious sensitivities.
    \item Most of our neighbors do not meet our criterion for post-starburst and quiescent galaxies ($\mathrm{log_{10}(sSFR)}$ $<-9.8$ yr$^{-1}$), but are still actively forming stars, which is inconsistent with the conformity effect at low-$z$. We attribute this to the long timescale of conformity, which can lead to protoclusters not being as uniformly quiescent or star-forming as they would be several Gyr later at $z<1$.
    \item We also locate our S$\mathrm{\hat{O}}$NG sample in environments which are denser than the field but marginally less dense or equally dense with respect to the environments of mass-matched, coeval SFGs. This differs from the well-known low-$z$ trend of QGs residing in local overdensities and resembles recent modeling results \citep[e.g.,][]{DeLucia2025, Kimmig2025} where high-$z$ QGs can inhabit diverse environments other than local overdensities. Still, given the large uncertainties from our MC method, caution is warranted.
    \item Using our method of sectoral number density, we also identify traces of filamentary structures for a small percentage (4\%) of the S$\mathrm{\hat{O}}$NG QGs, the most remarkable being S$\mathrm{\hat{O}}$NG-133 with two distinctly oriented filament-like overdensities. This suggests that some of the QGs may rejuvenate over time by cold gas accretion. Although our method remains rudimentary, it enables preliminary identifications that could help guide spectroscopic follow-ups.
    \item Lastly, applying our procedure to a limited sample of spectroscopically confirmed 3 $<$ $z$ $\leq$ 4.5 QGs, we identify four separate protoclusters around four QGs. None of these systems show filamentary signals, and similar to the S$\mathrm{\hat{O}}$NG sample, they also do not support conformity. Saigon, an intermediate-mass QG at $z$ $=$ 4.55 identified in this work, is by far the most distant low-mass QG ever found (M$_*$ $= 1.33 \times10^9$ M$_\odot$), and one of the earliest spectroscopically confirmed QGs located in a protocluster node; it was likely quenched at $z$ $\sim$ 5 by starburst-induced gas consumption. SED fitting also shows that its neighbor HX8 has a mass ratio $\sim$ 3:1, suggesting Saigon may not be the protocluster central. High-resolution, long-exposure spectroscopic follow-ups are needed to tackle these uncertainties.
    
\end{itemize}

\section{ACKNOWLEDGMENTS}

This work is dedicated to Vietnamese astrophysicists, who have persisted in exploring the cosmos and inspiring the youth despite the scarcity of resources and opportunities. Many formalisms in this paper are derived from the Vietnamese language: the suffix HX stands for \textit{hàng xóm} (neighbor); the sector count $q$ originates from \textit{quạt} (fan), the technical term for a circular sector.

N.B. would like to thank their elderly parents Nguyễn Hoà and Nguyễn Thị Oanh, who grew up identifying the dark sky with deadly airstrikes and knew little of astrophysics, but still let their child move across the globe to pursue their dream. N.B. would also like to thank their sisters Nguyễn Kim Ngọc and Nguyễn Phương Anh for inspiring their higher-education journey and supporting them financially during their undergraduate years.

This work is based mainly on observations made with the NASA/ESA/CSA JWST. The data were obtained from the Mikulski Archive for Space Telescopes at the Space Telescope Science Institute, which is operated by the Association of Universities for Research in Astronomy, Inc., under NASA contract NAS 5-03127 for JWST. Some of the data products presented herein were retrieved from the DAWN JWST Archive (DJA). DJA is an initiative of the Cosmic Dawn Center (DAWN), which is funded by the Danish National Research Foundation under grant DNRF140.

This project has received funding from the European Union’s Horizon 2020 research and innovation program under the Marie Sklodowska-Curie grant agreement No 101148925. French COSMOS team members are partly supported by the Centre National d’Etudes Spatiales (CNES). O.I. acknowledges the funding of the French Agence Nationale de la Recherche for the project iMAGE (grant ANR-22-CE31-0007). L.M. acknowledges the financial contribution from the PRIN-MUR 2022 20227RNLY3 grant “The concordance cosmological model: stress-tests with galaxy clusters” supported by Next Generation EU and from the grant ASI n. 2024-10-HH.0 “Attività scientifiche per la missione Euclid – fase E”. G.E.M. acknowledges the Villum Fonden research grants 37440 and 13160. O.R.C. is supported by an NSF Astronomy and Astrophysics Postdoctoral Fellowship under award AST-2503202. N.E.D acknowledges support from NSF grants LEAPS-2532703 and AST-2510993.

This work has used the following software packages: NumPy \citep{numpy}, matplotlib \citep{matplotlib}, Astropy \citep{astropy1, astropy2, astropy3}, pandas \citep{pandas}, SciPy \citep{scipy}, and seaborn \citep{seaborn}. Additionally, this research made use of Regions, an Astropy package for region
handling \citep{regions}. Lastly, this work is partly based on tools and data products produced by GAZPAR operated by CeSAM-LAM and IAP.

\appendix

\section{Additional Figures} \label{spectra_appendix}

\begin{figure}[b]
	\includegraphics[width=\textwidth]{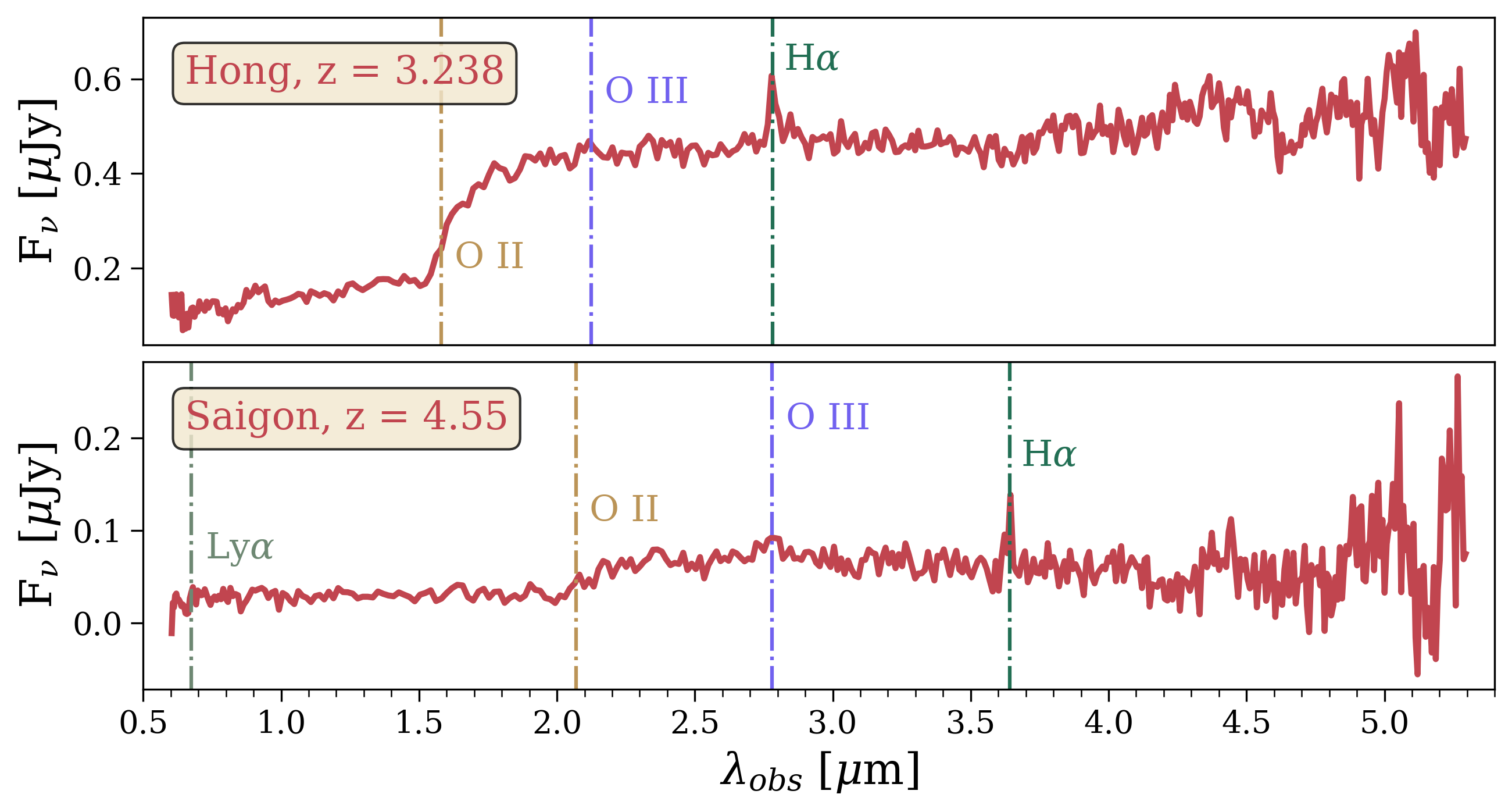}
    \caption{NIRSpec/PRISM spectra of Hong (top) and Saigon (bottom). All wavelengths are in the observed frame.}
    \label{fig:two_spectra}
\end{figure}

\bibliography{sparsebytheriver}{}
\bibliographystyle{aasjournalv7}

\end{document}